\documentclass[showpacs, prb,twocolumn,preprintnumbers , superscriptaddress, aps]{revtex4-2}
 
\usepackage{color}
\usepackage{amsmath,amssymb}
\usepackage{pifont}
\usepackage{amssymb}  
\usepackage{bbold}
\usepackage{float}
\usepackage{subfloat}
\usepackage{amsmath,amssymb}
\usepackage{mathtools}
\usepackage{physics}
\usepackage{breqn} 

\usepackage[caption=false]{subfig}
\usepackage{tikz}
\usepackage{makecell}
\usepackage{subfig}
\usepackage{pifont}   
\usepackage{graphicx} 
\usepackage{dcolumn}  
\usepackage{bm}       
\usepackage{multirow} 
\usepackage{placeins}
\usepackage[colorlinks]{hyperref}
\usepackage{mathtools}

\usepackage{breqn}
\usepackage{mathrsfs}

\usepackage{multirow}

\newcommand\scalemath[2]{\scalebox{#1}{\mbox{\ensuremath{\displaystyle #2}}}}

\begin{document}

\title{Analytical Treatment of Noise-Suppressed Klein Tunneling in Graphene with Possible Implications for Quantum-Dot Qubits}

    \author{Kamal Azaidaoui}
    \affiliation{Laboratory of Theoretical Physics$,$ Faculty of Sciences$,$ Choua\"ib Doukkali University$,$ PO Box 20$,$ 24000 El Jadida$,$ Morocco}
		
    \author{Ahmed Jellal}
		\affiliation{Laboratory of Theoretical Physics$,$ Faculty of Sciences$,$ Choua\"ib Doukkali University$,$ PO Box 20$,$ 24000 El Jadida$,$ Morocco}
        \author{Hocine Bahlouli}
		\affiliation{Physics Department and IRC Advanced Materials$,$
		King Fahd University
		of Petroleum $\&$ Minerals$,$
		Dhahran 31261$,$ Saudi Arabia}
    \author{A. Al Luhaibi}
        \affiliation{Physics Department$,$
		King Fahd University
		of Petroleum $\&$ Minerals$,$
		Dhahran 31261$,$ Saudi Arabia}
      \affiliation{Interdisciplinary Research Center (IRC) Quantum Center$,$ KFUPM$,$ Dhahran$,$ Saudi Arabia}
    \author{Michael Vogl}
		\affiliation{Physics Department$,$
		King Fahd University
		of Petroleum $\&$ Minerals$,$
		Dhahran 31261$,$ Saudi Arabia}
      \affiliation{Interdisciplinary Research Center (IRC) Quantum Center$,$ KFUPM$,$ Dhahran$,$ Saudi Arabia}

\begin{abstract}
  We study quantum tunneling through a potential barrier whose height fluctuates in time and is modeled by Gaussian white noise. We map the stochastic dynamics onto an equivalent time-independent Lindblad equation for the density matrix, allowing fully analytical solutions. For  non-relativistic particles as described by the Schr\"odinger equation,  noise introduces dissipation that suppresses Fabry-P\'erot oscillations and yields an exponentially decaying transmission. Applying the same formalism to graphene, we demonstrate that noise induces a complex longitudinal wavevector within the barrier, leading to a strong suppression of transmission and Klein tunneling, even at normal incidence. Our approach promises improved control over Klein tunneling. These results demonstrate that noisy barriers can act as tunable dissipative elements, offering a pathway to enhanced control of electron transport in graphene-based devices. We also briefly discuss how our results could guide the design of graphene quantum dots for potential use in spin qubit devices.
  
\end{abstract}

\maketitle

\section{Introduction}

Graphene is a monolayer of carbon atoms arranged in a honeycomb lattice. It was successfully isolated in 2004 as the first stable $2D$ crystal \cite{Novoselov2004}. Its discovery was a surprise because its existence contradicted theoretical expectations that fluctuations would destroy long-range order in 2D systems \cite{Mermin1968,AshcroftMermin1976}. Its discovery led to intense research activity for two-dimensional electronic systems \cite{Novoselov2004,Geim2009}. At low energies, electrons in graphene behave as massless Dirac fermions, leading to exotic transport phenomena, like the half-integer quantum Hall effect \cite{Novoselov2005,CastroNeto2009}. Graphene's unusual electronic properties, together with the exceptional mechanical and thermal properties, have turned graphene into one of the most widely studied systems in condensed matter physics and material science \cite{CastroNeto2009,Geim2009}.

 Graphene has attracted considerable interest because of its remarkable potential for technological applications. In particular, its exceptionally high carrier mobility and large saturation drift velocity under strong electric fields make it a promising material for high-frequency graphene field-effect transistors and radio-frequency electronics \cite{Schwierz2010, Petrone2013, Dorgan2010}.
In addition, graphene combines mechanical flexibility with optical transparency, making it attractive for flexible, transparent conductors and devices \cite{Kim2010,Torrisi2012,Georgiou2013,electronics12010045}.
Another interesting physical effect in graphene, with great potential for technological applications but also with possible drawbacks, is the so-called Klein tunneling \cite{Katsnelson2006, CastroNeto2009}. Specifically, in graphene, pseudospin is locked to carrier momentum. For an ordinary potential barrier (e.g., purely electrostatic potential), chirality at normal incidence is conserved, and backscattering is therefore forbidden. As a result, an electron incident perpendicular to the barrier can transmit with unit probability, independent of the barrier height and width. This effect has inspired early works on low-resistance ballistic transport and electronic optics in gate-defined junctions \cite{Beenakker2008,CastroNeto2009}. Moreover, it demonstrates graphene's potential for high-speed electronics and transistor devices \cite{Schwierz2010}.

However, Klein tunneling also has drawbacks for device control, as it makes it difficult to control electrons with electrostatic barriers. In particular, the normally incident channel cannot be switched off by increasing the height or width of the barrier. As a result, for these types of barriers in graphene, one struggles to achieve a large on/off ratio, which is a key requirement for transistor applications \cite{Schwierz2010,CastroNeto2009}.

To gain control over Klein tunneling, several strategies have been proposed. One of these strategies uses gate-defined $p$--$n$ and $p$--$n$--$p$ junctions which act as optoelectronic elements, where the refraction at a $p$--$n$ interface can collimate ballistic carriers \cite{Cheianov2007},  and $p$--$n$--$p$ cavities exhibit interference patterns which are sensitive to angle and density \cite{Young2009,Stander2009}. Other works suggest using periodic potentials or appropriate substrates to reshape the Dirac dispersion and generate additional Dirac points, thereby strongly modifying the angular dependence of transmission \cite{Park2008, Park2008b}. In addition, magnetic barriers offer a distinct control mechanism that can confine Dirac fermions and produce strongly direction-dependent tunneling \cite{DeMartino2007,Masir2009}. In more engineered geometries with a band gap (for instance, induced by an appropriate substrate) or combined electrostatic and magnetic barriers, it is possible to adjust transmission and related beam-shift effects \cite{Mekkaoui2019}. More recently, Klein tunneling and related transport phenomena have continued to be investigated in increasingly realistic graphene-based systems and engineered Dirac materials, highlighting the sustained interest in controlling Dirac-fermion transport \cite{PhysRevLett.132.146302,Tran_2024,Liu2023,Marin-Colli2026}.

The strategies we discussed above are based on static (time-independent) barriers, meaning they rely on time-independent control mechanisms that result in unitary dynamics and energy conservation throughout evolution. Moreover, even though they can reshape the transmission by modifying the matching conditions at the barrier interfaces and offer angle- or energy-selective filtering, they cannot open inelastic channels or introduce controlled loss or gain mechanisms. Furthermore, this limitation has been overcome in graphene, where the normally incident mode is protected by chirality in an elastic scattering problem, leading to the absence of robust switch-off by a purely electrostatic barrier. Therefore, examining this problem in the time domain may expand the scope of this control. When the barrier becomes time-dependent and the scattering becomes inelastic due to the exchange of carriers' energy with the drive, additional channels can open. This regime has been extensively studied for periodically driven barriers through the Tien–Gordon (photon-assisted tunneling) mechanism, where an ac gate opens Floquet sidebands while coherent transport remains unitary \cite{PhysRevB.75.035305,PhysRevB.87.125422}. In contrast, the present work considers stochastic Gaussian white-noise modulation, which, after averaging over noise, leads to an effective Lindblad description with dissipation rather than coherent sideband formation.
 
 A well-known approach to controlling quantum devices in the time domain is Floquet engineering. Here, system parameters are modulated periodically in time \cite{Eckardt2017,Bukov2015}. This periodic driving can effectively reshape band structures and open gaps in ways that are hard to achieve with equilibrium strategies \cite{Eckardt2017,Bukov2015}. A famous example is graphene subjected to circularly polarized light, which induces a non-zero gap and generates a Hall response even in the absence of an external magnetic field \cite{OkaAoki2009}. Periodic driving has been proposed as a knob to renormalize interlayer couplings and thereby tune the effective magic-angle physics of twisted bilayer graphene \cite{Vogl2020}. More broadly, our work belongs to the growing field of time-dependent quantum transport, where periodic driving, pumping, and other nonequilibrium protocols are used to control electronic transport; for a recent overview, see Ref.~\cite{Acciai2025}.

Although Floquet driving offers a powerful way to design effective band structures, it still imposes strict constraints where the dynamics are structured around a single driving frequency and its harmonics. Another interesting time domain protocol is to include random temporal modifications -noise. In realistic graphene devices, potential fluctuations may arise from several physical sources. Charge impurities or trap states in the substrate can randomly capture and release electrons, producing time-dependent Coulomb potentials that act on electrons in the graphene layer. In addition, electromagnetic noise in the measurement circuitry can cause fluctuations in the gate voltage controlling the barrier potential. These effects generate time-dependent electrostatic potentials that can be modeled as stochastic fluctuations of the barrier height.

Noise offers additional freedom because it depends on a wider range of time scales (unlike Floquet systems, there is no single period that sets a time scale). It can be used to bridge the gap between coherent control and decoherence/dissipation control. The idea that dissipation can qualitatively modify quantum tunneling is well established in the Caldeira–Leggett theory of dissipative tunneling and in the related dissipative two-state problem \cite{PhysRevLett.46.211,CALDEIRA1983374,RevModPhys.59.1,doi:10.1142/8334}. Such ideas have also been described in the context of driven transport and open quantum systems \cite{Kohler2005,BreuerPetruccione2002}. It has been shown that time-dependent gate barriers in graphene generate transport properties that are not present in static ones \cite{Savelev2012}, suggesting that the Dirac tunnel can be highly sensitive to the temporal structure  \cite{Savelev2012}. Therefore, we are motivated to focus on a new driving mechanism beyond Floquet theory - noise as a stochastic,  engineered modulation. This driving provides a simple way to incorporate dissipative effects into an analytically solvable barrier-scattering problem.
Among possible stochastic processes, Gaussian white noise is a very convenient starting point because it is the minimal broadband process and fully defined by its initial moments. Importantly, its delta correlations permit a mapping to an effective time-independent Lindblad equation \cite{Chenu2017,PhysRevB.109.L220406}. This observation is essential to our work because, as we will see, it enables a fully analytical treatment.
Once we have established a description of the density matrix for a fluctuating barrier, we apply our formalism to graphene to reexamine Klein tunneling \cite{Katsnelson2006,Beenakker2008}. Our main question is whether a temporal noisy barrier can suppress this ideal conduction channel. This goal can be achieved by computing the transmission coefficient as a function of the incidence angle in the presence of noise.

Our paper is structured as follows. In Sec.~\ref{sec:model}, we introduce the setup of the tunneling problem. In particular, we define the random-noise barrier for both non-relativistic particles as a benchmark and for massless Dirac fermions in graphene. In Sec.\ref{sec:time_indep}, we average over noise configurations and map the stochastic dynamics onto an effective, time-independent Lindblad equation. This approach allows us to reformulate our problem as a stationary scattering problem in the $(x,x')$ plane - coordinates have doubled compared to the wavefunction approach because density matrices depend on two sets of coordinates. In Sec.~\ref{sec:results}, we fully analytically solve our problem and present results for transmission, reflection, and absorption probabilities (absorption in the sense of missing coherent flux). We first study the 1D Schr\"odinger case and then apply the same formalism to graphene to demonstrate that Klein tunneling is suppressed. Next, in Sec. \ref{sec:App}, we briefly discuss how our results may inform quantum dot design for spin qubits in graphene. Finally,  Sec.~\ref{sec:conclusion} summarizes our main conclusions and discusses their implications and possible extensions. Appendices contain the technical derivations and explicit expressions for the density matrix, along with details on the matching conditions and coefficients.
\section{Model and problem setup}
\label{sec:model}
Quantum tunneling through time-independent barriers like single electrostatic and magnetic barriers as well as periodic superlattice potentials has been investigated extensively in mesoscopic transport and, in graphene in particular, where the chirality gives rise to Klein tunneling and Fabry-P\'erot interference \cite{Beenakker2008, CastroNeto2009, Katsnelson2006, DeMartino2007, Masir2009, Park2008, Park2008b, Barbier2010}.
Simultaneously, tunneling through time-periodically driven barriers, which leads to photon-assisted (Floquet) transport, has a longstanding history in transport theory and has also been explored in graphene-based contexts \cite{TienGordon1963, PedersenButtiker1998, PlateroAguado2004, Kohler2005, Savelev2012, Jellal2014}. Therefore, in this work, we consider a different type of time dependence to study tunneling across noisy barriers. That is, we consider barriers that fluctuate stochastically over time and investigate how this effect modifies transport properties. In particular, we investigate how such temporal noise can be used as a useful technique to control transmission rather than merely an unfavorable source of decoherence.

\begin{figure}[ht!]
    \centering
\includegraphics[width=\columnwidth]{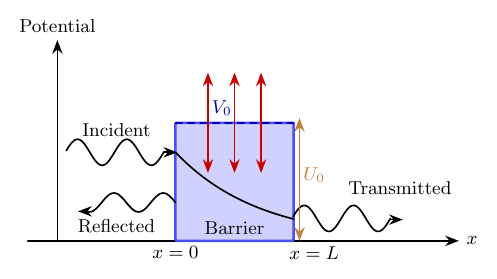}
    \caption {Quantum tunneling across a barrier of static height $U_0$, whose amplitude fluctuates in time with noise strength $V_0$. The temporal fluctuations introduced by noise are indicated by the red arrows. Moreover, the incident, reflected, and transmitted waves are schematically indicated.}
    \label{fig:noisy_barrier}
\end{figure}

Here, Fig.~\ref{fig:noisy_barrier} illustrates the schematic setup we had in mind that describes the tunneling across a one-dimensional noisy barrier, where its height fluctuates in time around an average value $V_0$.
 To study tunneling across a noisy barrier, we set up a stochastic single-particle Hamiltonian as
\begin{align}
\hat{H}_{St}(t) = \hat{H}_T + V(\hat{x},t),
\label{eq:Hst_general}
\end{align}
where $\hat{H}_T$ is the Hamiltonian of the target system, and $V(\hat{x},t)$ is the time-dependent fluctuating barrier. In the following, we identify $\hat{H}_T$ for the two systems of interest.
\subsection{Massless Dirac fermions in graphene.}
We first consider low-energy charge carriers in single-layer graphene, which are governed by the Dirac-Weyl Hamiltonian
\begin{align}
\hat{H}_T^{(D)} =  v_F \bm{\sigma}\cdot\hat{\mathbf{k}}+ U(\hat{x})\mathbf{I}_2,
\label{eq:HT_graphene}
\end{align}
where $\hat{\mathbf{k}}=(-i\partial_x,-i\partial_y)$ is the momentum operator, 
$v_F= 10^6\,\mathrm{m/s}$ is the Fermi velocity (we chose units such that $\hbar=1$), 
$\bm{\sigma}=(\sigma_x,\sigma_y)$ is a vector of Pauli matrices, 
and 
\begin{equation}
    U(x) =
    \begin{cases}
        U_0, & 0 < x < L,\\
        0,   & \text{otherwise}.
    \end{cases}
    \label{eq:U_rectangular}
\end{equation}
 is a static rectangular barrier term.
\subsection{Non-relativistic particles}
For comparison, and to address the question of whether the observed effects are generic or specific to the system under consideration, we next examine the case of a non-relativistic particle of mass $m$, which is described by

\begin{align}
\hat{H}_T^{(S)} = \frac{\hat{p}_x^2}{2m} + U(\hat{x}).
\label{eq:HT_sch}
\end{align}
with $\hat{p}_x=-i\hbar\partial_x$ as the momentum operator along the transport direction.

\subsection{Explicit noisy barrier term}
\begin{figure}[ht!]
    \centering
\includegraphics[width=\columnwidth]{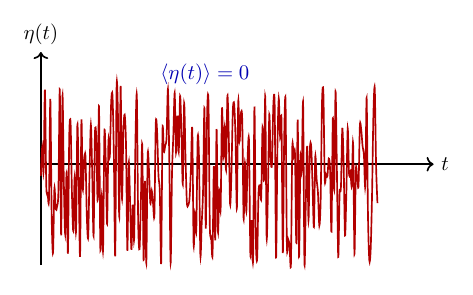}
    \caption{Sample realization of Gaussian white noise fluctuations, the noise has zero mean and shows $\delta$ temporal correlations, $\langle\eta(t)\eta(t')\rangle=\delta(t-t')$.}
    \label{fig:Gaussian_Profile}
\end{figure}

The term $V(\hat{x},t)$, which results from the introduction of noise in our study, is being modeled as a product of a spatial operator and a stochastic strength.
\begin{equation}
V (\hat{x}, t) = \eta(t) V (\hat{x}),
\end{equation}
where $V(\hat{x})$ defines the noisy barrier strength (noise strength was absorbed into this quantity), which is defined by:
\begin{align}
V(x) = \begin{cases}
V_0, & 0 < x < L,\\
0,   & \text{otherwise}.
\end{cases}
\end{align}
Accordingly, the spatial profile of the barrier remains unchanged throughout this work, while only its height fluctuates in time about the static barrier.
Moreover, $\eta(t)$ represents a noise profile. An example of a
possible noise profile can be seen in Fig.~\ref{fig:Gaussian_Profile}.
We will now restrict our discussion to Gaussian white noise, which has attractive properties. In particular, it has a zero mean and is assumed to have a $\delta$-function time correlation. That is, it fulfills the following conditions
\begin{equation}
\langle \eta(t) \rangle = 0,
\qquad \langle \eta(t)\,\eta(t') \rangle = \delta(t-t').
\label{Gaussian_white_noise}
\end{equation}
We choose units such that $\hbar=1$ and normalize the Schr\"odinger equation by an arbitrary reference energy scale $E_0$. Equivalently, time is measured in units of $\hbar/E_0$, so the variable $t$ in Eq.~(\ref{Gaussian_white_noise}) is dimensionless, and therefore $\delta(t-t')$ is dimensionless. Often in what follows, we use $U_0$ as our reference energy scale.
The time correlation $\langle \eta(t)\,\eta(t') \rangle = \delta(t-t')$ ensures a memoryless or Markovian nature of dynamics, while Gaussianity guarantees that the noise is completely characterized by its first two moments.

The white-noise approximation should be understood as an idealization corresponding to the regime in which the correlation time of the barrier fluctuations is much shorter than the characteristic time an electron spends traversing the barrier. In this limit, the electron effectively experiences delta-correlated fluctuations, and the dynamics become Markovian. Although no experimental noise source is perfectly delta-correlated, this approximation provides an effective description whenever the relevant noise spectrum is sufficiently broad over the transport frequencies.

Despite these nice simplifying features of Gaussian white noise, solving the time-dependent Schr\"odinger equation for the stochastic Hamiltonian $\hat{H}_{St}(t)$ is computationally expensive. One reason is that time evolution by itself is more costly than solving a static equation \cite{PlateroAguado2004, Kohler2005}. Second, one would need to do an average over a large number of noise realizations to obtain physically meaningful results \cite{Dalibard1992, Molmer1993, PlenioKnight1998, BreuerPetruccione2002}. Therefore, in the next section, we circumvent this burden by exploiting the relationship between classical Gaussian white noise and the static Lindblad equation, thereby recasting the problem in an effectively time-independent form.
\section{Time independent formalism}
\label{sec:time_indep}

Next, our task will be to average over all possible noise configurations, which, as shown in \cite{Chenu2017} provides an exact mapping from the time-dependent Schr\"odinger equation to an effective time-independent Lindblad equation. The important point to notice is that while individual noise realizations are governed by a stochastic Schr\"odinger equation which maintains state purity and is governed by unitarity evolution, the noise-average density matrix $\rho(t)=\langle\ket{\Psi_{\eta}(t)}\bra{\Psi_{\eta}(t)}\rangle$, evolves according to deterministic, non-unitary dynamics which cannot be described by Hamilton's evolution alone. More precisely, for Gaussian white noise, it will turn out that a Lindblad equation will be needed to properly describe dynamics, while other colors of noise lead to more complex dynamics that mimic a system connected to a non-Markovian bath \cite{Chenu2017, BreuerPetruccione2002}.\\
To understand the link to Lindblad dynamics, we first establish a general mapping by starting from the stochastic Hamiltonian defined in Eq.~(\ref{eq:Hst_general}), where $\eta(t)$ satisfies the correlation relations indicated in Eq.~(\ref{Gaussian_white_noise}). Then the evolution of the stochastic density matrix can be expressed as 
\begin{align}
  \frac{d\rho_\eta(t)}{dt}
  = -i[\hat{H}_T, \rho_\eta(t)]\
   - i\eta(t) \left[V(\hat{x}), \rho_\eta(t)\right].
\end{align}
Averaging over all noise configurations and applying Novikov's theorem for Gaussian white noise \cite{Chenu2017,PhysRevB.109.L220406}, one obtains a Markovian master equation of Lindblad form \cite{10.1063/1.522979,Lindblad1976}
\begin{equation}
  \frac{d\rho}{dt}= -i[\hat{H}_T, \rho]  + V(\hat{x}) \rho V(\hat{x}) - \frac{1}{2} \{ V(\hat{x})^2, \rho \}.
  \label{Lindblad_Eq}
\end{equation}
where $\rho(t) = \langle \rho_\eta(t) \rangle$ is the noise averaged density matrix.
We emphasize that this construction differs from the Caldeira–Leggett bath model. In the Caldeira–Leggett formulation, dissipation arises by integrating out a quantum environment and the result depends on the bath spectral density \cite{CALDEIRA1983374,doi:10.1142/8334}. Here, in contrast, the barrier height is driven by a classical Gaussian white-noise process, and the noise average produces a local-in-time Lindblad term with jump operator $V(\hat{x})$. The resulting Lindblad equation Eq~(\ref{Lindblad_Eq}) clearly demonstrates that we have transformed a very complex stochastic problem into a simple time-independent deterministic equation, which, as we will see later for our case, can be solved analytically.
In both cases, we want to consider the Lindblad equation, which can be written explicitly.\\
First, for the 1D non-relativistic Schr\"odinger equation case, we obtain 
\begin{equation}
\scalemath{0.9}{\frac{d\rho}{dt} = -i\Big[\frac{\hat p_x^2}{2m}+U(\hat x),\,\rho\Big]
+\left(V(\hat{x})\,\rho\,V(\hat{x})-\tfrac12\{V(\hat{x})^2,\rho\}\right)}.
\label{eq:L_Sch}
\end{equation}
And secondly, for the case of graphene at low energy, we obtain
\begin{equation}
\scalemath{0.8}{\frac{d\rho}{dt} = -i\Big[ v_F\,\boldsymbol{\sigma}\cdot\hat{\mathbf k}+U(\hat x)\,\mathbf{I}_2,\,\rho\Big]
+\left(V(\hat{x})\,\rho\,V(\hat{x})-\tfrac12\{V(\hat{x})^2,\rho\}\right).}
\label{eq:L_Dirac}
\end{equation}
We note that both results are, so far, written abstractly, without us having chosen an explicit basis for the density matrix. To render the equations into their most useful form, we now switch to a position basis and define
\begin{align}
\rho(x,x')= \langle x|\rho |x'\rangle.
\end{align}

\begin{figure}[ht!]
    \centering    \includegraphics[width=\columnwidth]{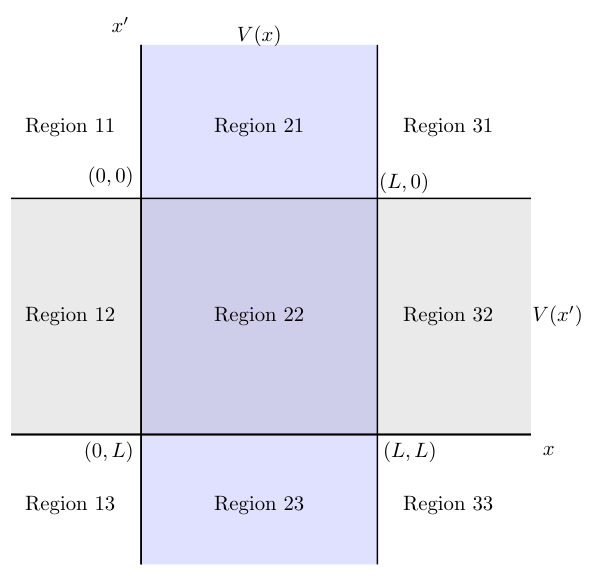}
    \caption{Schematic of the nine regions in the $(x,x')$ plane used in the density-matrix scattering formalism. The vertical and horizontal lines at $x=0$ and $ x=L$, and at $ x'=0$ and $ x'=L$, mark the interfaces of the noisy barrier. Each region $(i,j)$ corresponds to $x$ and $x'$ being in the left ($x<0$), inside the barrier, ($0<x<L$), or in the right lead ($x>L$).}
    \label{fig:nine_regions}
\end{figure}

We observe directly that the formalism of the density matrices differs considerably from the wavefunction approach, in that it includes two separate spatial variables - position space is, in a sense, doubled. \\
Much like the square-barrier case for wavefunctions, we observe that the equation can be split into various spatially uniform regions ($U(x)$ and $V(x)$, we recall, are piecewise constant) - we need to consider both copies of spatial coordinates separately. In Fig. \ref{fig:nine_regions}, we see the nine different regions one ends up with.

Importantly, this observation means that, for density matrices, one must match the boundaries between the nine regions. In the 1D Schr\"odinger case, one has to match wavefunctions and their derivatives (it is a differential equation with second-order derivatives). In the case of graphene, which is first-order in derivatives, one only needs to match the different components of the wavefunction. In the non-noisy case, this seems unusually challenging compared to the typical wavefunction approach, which only requires matching three regions. However, we are aided by the fact that the density matrix is hermitian: $\rho^\dagger = \rho$. Importantly, this property imposes useful constraints on the number of allowed solutions. Such constraints significantly reduce the complexity of our nine-region matching problem. Most useful for our purposes, the condition can be expressed in position space as 
\begin{align}
\rho_{ij}^\dagger(x, x^{\prime}) = \rho_{ji}(x^{\prime},x),
\label{eq:her_cond}
\end{align}
where, $i,j\in\{1,2,3\}$, indicate the different regions shown in Fig. \ref{fig:nine_regions}.
Because we deal with density matrices rather than wavefunctions, we of course also need to be careful in how we determine transmission and reflection coefficients. Often, in the wavefunction approach, we have wavefunctions like (for a simplest example)
\begin{equation}
    \psi_{in}=e^{ikx}+re^{-ikx};\quad \psi_{out}=te^{ikx}
\end{equation}

and one interprets the tunneling coefficients $t$ and $r$ of different propagating mode contributions accordingly - this is a shortcut. However, for the density matrix, we do not know directly how transmission and reflection relate to coefficients. Therefore, a reorientation to basics is necessary. In particular, we start from a more physical idea of currents across various spatial regions. Since currents can be computed as the expectation value of current operators $J_i$ via
\begin{equation}
    \langle J_i\rangle=\mathrm{Tr}(J_i\rho)
    \label{eq: current_expectations}
\end{equation}
They provide a natural language for density matrices.\\
Here, we may use probability current operators for the Schr\"odinger and Dirac cases. where the transmission and reflection probabilities are 
\begin{align}
T = \frac{\langle J_x \rangle_{\text{tr}}}{\langle J_x \rangle_{\text{in}}},\quad
R = \frac{\langle J_x \rangle_{\text{re}}}{\langle J_x \rangle_{\text{in}}}.
\label{eq:Tunneling_probabilities}
\end{align}

Here, $\langle J_x\rangle_{\text{in}}$ is the incident current in the left lead ($x<0$), evaluated far from the barrier. Similarly, $\langle J_x\rangle_{\text{re}}$ is the reflected current in the left lead, and $\langle J_x\rangle_{\text{tr}}$ is the transmitted current in the right lead ($x>L$), both evaluated in the asymptotic regions.
With our discussion now, all the pieces are in place to analytically solve the problem of noisy barriers.
\section{Results}
\label{sec:results}

After setting the theoretical framework, we are now ready to find analytical expressions for quantum tunneling through noisy potential barriers. We emphasize that the following analytical treatment of the tunneling problem is only made possible due to a reformulation in terms of a stochastic Schr\"odinger equation - using a time-independent deterministic density matrix approach instead. This description shift enables the derivation of closed-form expressions for transmission and reflection probabilities that capture the fundamental physics of noise-induced effects in quantum transport. Here, we will study two cases: non-relativistic particles in 1D and massless Dirac fermions in graphene. We start with the 1D Schr\"odinger Hamiltonian, both to gain simple insight and as a verification step. We then move to graphene to observe noise-suppressed Klein tunneling.

\subsection{Results for the non-relativistic particle}

For the non-relativistic particle,  we obtain the global solution of the density matrix in the nine regions as follows. First, we express Eq.~(\ref{eq:L_Sch}) in the position basis. Here, we observe that the Lindbladian is stationary such that we may choose a density matrix that evolves according to $\rho(t)=\rho_0 e^{-i(\epsilon-\epsilon')t}$. Our Lindblad equation then reduces to the stationary Lindblad equation given in Eq. \eqref{eq:rho_schrodinger_position}.

\begin{widetext}
\begin{align}
(\epsilon-\epsilon')\,\rho(x,x') =&
\left(-\frac{1}{2m}\partial_x^{2}+U(x)-i\frac{V^2(x)}{2}\right)\rho(x,x')-\rho(x,x')
\left(-\frac{1}{2m}\partial_{x'}^{2}+U(x')+i\frac{V^2(x')}{2}\right)\nonumber\\
&+iV(x)\rho(x,x')V(x').
\label{eq:rho_schrodinger_position}
\end{align}
\end{widetext}
We note that in region I (and III) the noise term vanishes, so the incident and transmitted plane waves have real energy $\epsilon$. Constants $(\epsilon-\epsilon')$ appearing in the stationary Liouville-space equation should be understood as spectral parameters; in the dissipative region they may effectively become complex, encoding decoherence rates.

We note that the potential $V(x)$ and $V(x^\prime)$ in Fig.~\ref{fig:nine_regions} are piecewise constants and depend on $x$ and $x'$. Therefore, in each region, the resulting  Eq.~(\ref{eq:rho_schrodinger_position}) is a linear equation with constant coefficients. Hence, we can find solutions for the density matrix that can be written as a tensor product for each region  $\rho_{ij}(x,x')=\rho_{i}(x)\otimes \rho_j^{\dagger}(x')$. Substitution of this ansatz reduces the problem to two independent stationary equations, one that governs the motion of a wave function $\rho_i(x)$  and the second for the associated wave function $\rho^{\dagger}_j(x')$.

Solving the time-independent Lindblad equation (~\ref{eq:rho_schrodinger_position}) in all nine regions of Fig.~\ref{fig:nine_regions} allows us to determine all elements of the stationary density matrix $\rho_{ij}(x,x')$. The hermiticity condition of the density matrix $\rho=\rho^\dagger$ (see also Eq. \eqref{eq:her_cond}) can directly be used to reduce the number of independent coefficients when setting up linear combinations of plane waves for the different regions. An explicit form of $\rho_{ij}(x,x')$ in all nine regions is relegated to Appendix~\ref{app:sch_rho_nine_regions} for brevity.

We obtain the tunneling coefficients for this case of the non-relativistic particle by imposing continuity conditions at the input $(0,0)$ and output $(L, L)$ faces in the plane $(x,x')$. Because the $H_T^{(S)}$ is a second-order derivative in position space, the wavefunctions must be continuous together with their derivatives. More details are given in the Appendix \ref{app:sch_matching}. 

With the general solution of the density matrix for all regions, we are ready to extract tunneling probabilities. For this purpose, here, we need to introduce the probability current. For the case of the non-relativistic particle, the current operator along the tunneling directions is
\begin{align}
\hat{J}_x = \frac{e}{2m} \left[ |x\rangle\langle x|\hat{p} + \hat{p}|x\rangle\langle x| \right]
\end{align}
We plugged this expression into Eq.~(\ref{eq: current_expectations}), which allows us to compute the probability currents from the density matrix expression in the position space:
\begin{align}
\langle J_x \rangle = \frac{e}{2mi}(\partial_x - \partial_{x^{\prime}}) \rho(x, x^{\prime})\big|_{x=x'}
\end{align}
Therefore, the expressions of probability currents at the leads are
\begin{align} 
		&\langle J_{x}\rangle_{\text{in}}=\frac{e k  }{m},\\
        &\langle J_{x}\rangle_{\text{re}} =-\frac{e k }{m}rr^{*},\\
       & \langle J_{x}\rangle_{\text{tr}}=\frac{e k }{m}tt^{*}.
    \label{eq:current_schro}
	\end{align}

Hence, the expressions for the probability currents allowed us to obtain those for the transmission and reflection probabilities.
\begin{align}
&T=\frac{\langle J_x \rangle_{tr}}{\langle J_x \rangle_{in}}=tt^{*}
,\qquad
R =\frac{\langle J_x \rangle_{re}}{\langle J_x \rangle_{in}}=rr^{*}.
\end{align}
By using the explicit expressions of the tunneling amplitudes $t$ and $r$, given in Appendix~\ref{app:sch_matching}, we compute the tunneling probabilities of the non-relativistic particle 
\begin{align}
&T=\frac{4k^2(q_1^2 + q_2^2)}{D_0 + D_1 e^{2q_2L} + D_2 e^{-2q_2L}},
\label{eq:T_sch_noise}
\\
&
R = \frac{N_0(\cosh(2q_2L)-\cos(2q_1L))}{D_0 + D_1 e^{2q_2L} + D_2 e^{-2q_2L}}.
\label{eq:R_sch_noise}
\end{align}
Here, $k$ is the wavevector at the leads, and $q=q_1+iq_2$ is the wavevector inside the barrier region, and their explicit expressions are 
\begin{align}
k &= \sqrt{2m\epsilon},
\label{eq:k_leads_sch}\\[2pt]
q_1 &=
\left[
m(\epsilon-U_0)
+\sqrt{
\left(m(\epsilon-U_0)\right)^2
+\left(\frac{mV_0^2}{2}\right)^2
}
\right]^{1/2},
\label{eq:q1_sch}\\[2pt]
q_2 &= \frac{mV_0^2}{2q_1}.
\label{eq:q2_sch}
\end{align}
With, the coefficients $D_0$, $D_1$, $D_2$, and $N_0$ are real functions of the system parameters and given as
\begin{widetext}
\begin{align}
&	D_0=-2 k q_2 \left(k^2-q_1^2-q_2^2\right) \sin (2 L q_1)-\frac{1}{2}\left(k^4-2 k^2 \left(q_1^2+3 q_2^2\right)+\left(q_1^2+q_2^2\right)^2\right) \cos (2 L q_1), \\
&D_{1,2} = \frac14\left[(k\pm q_1)^2+q_2^2\right]^2, 
\\
&
N_0 =\frac{1}{2} \left(k^4+2 k^2 \left(q_2^2-q_1^2\right)+\left(q_1^2+q_2^2\right)^2\right)
\end{align}
\end{widetext}
For completeness, Appendix~\ref{app:sch_matching} summarizes the matching procedure that leads to these expressions.

As explained earlier, tunneling in the presence of noise induces dissipation; a part of the incoming flux is absorbed (absorption in the sense of missing coherent flux), and the corresponding absorption probability is
\begin{align}
		A=1-T-R.
        \label{eq:A_sch_noise}
	\end{align}
Note that the Lindblad evolution preserves $\mathrm{Tr}(\rho)$
 and $A$ quantifies flux missing from the coherent plane-wave channels.\\
The effect of the white noise in the barrier is already visible at the level of the longitudinal wavevectors, where the particles in the leads propagate with a real wavevector $k$ [Eq.~\eqref{eq:k_leads_sch}]. Whereas as soon as they enter the noisy region, the longitudinal wavevector becomes complex $q=q_1+iq_2$ [Eqs.~\eqref{eq:q1_sch}--\eqref{eq:q2_sch}], where the condition of $q_2 > 0$ imposes an exponential decay and determines an attenuation length, where it measures the rate of decrease in probability of the flux during the passage of electrons through the noisy region. In contrast, the real part $q_1$ controls the usual oscillatory phase accumulation and therefore sets the Fabry-P\'erot interference scale inside the barrier.

Additionally, more physical impact of noise can be concluded from the analytical expression of the transmission probability in Eq.~(\ref{eq:T_sch_noise}), which is shown in the presence of the exponential factors $e^{\pm 2q_2 L}$ in the expression. In the limit of thick barriers  ($L\to \infty$), the dominant term is  $e^{ 2q_2 L}$, and  
\begin{equation}
T(L)\approx \frac{4k^{2}\,(q_{1}^{2} + q_{2}^{2})}{D_{1}}\, e^{-2 q_{2} L}.    
\end{equation}
Therefore, the transmission probability is exponentially damped with a decay rate proportional directly to $q_{2}$. Hence, barriers that are much larger than the damping length are effectively opaque $T(L\to\infty)=0$.

If we consider the limit for Eq.~\eqref{eq:R_sch_noise}  we find 
\begin{equation}
R(L\to\infty)\simeq \frac{N_{0}}{2 D_{1}} \leq 1.    
\end{equation}
 Therefore, the reflection probability saturates at a finite value. For this case, using Eq. \eqref{eq:A_sch_noise}, we find that the absorption probability
 \begin{equation}
     A(L\to\infty)=1-\frac{N_{0}}{2 D_{1}}
 \end{equation}
also saturates at a finite value. In conclusion, we find that the noise term acts as a lossy medium where only a fraction of the incoming flux is reflected - the missing part is absorbed.

In the absence of noise, $V_0=0$, the Lindblad master equation reduces to the von Neumann equation. In our problem, wave vectors become real-valued when $q_2=0$. This means that corresponding exponential factors in the expression for the tunneling probabilities disappear. In fact, we obtain well-known expressions for tunneling probabilities through a static rectangular barrier, and the dynamics are unitary with $R+T=1$.

\begin{figure}[ht!]
\centering 
\includegraphics[scale=0.42]{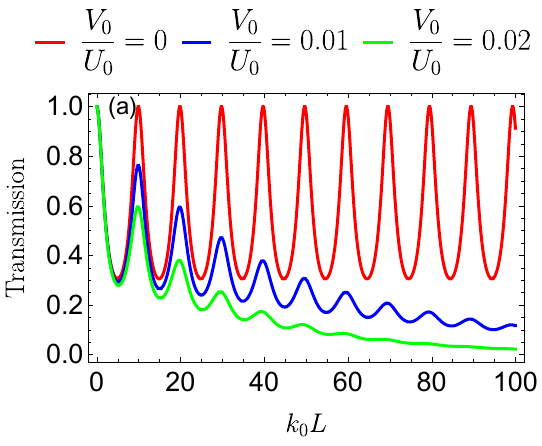}
\includegraphics[scale=0.42]{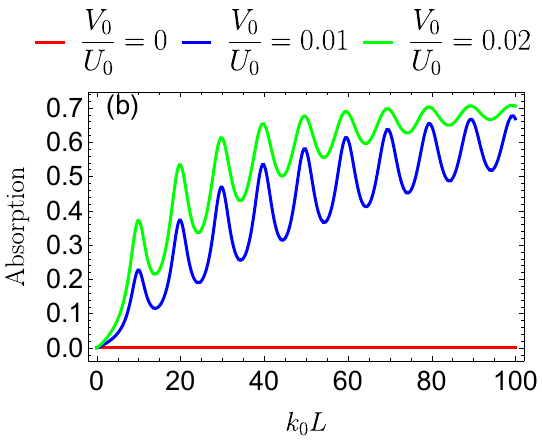}\\
\includegraphics[scale=0.42]{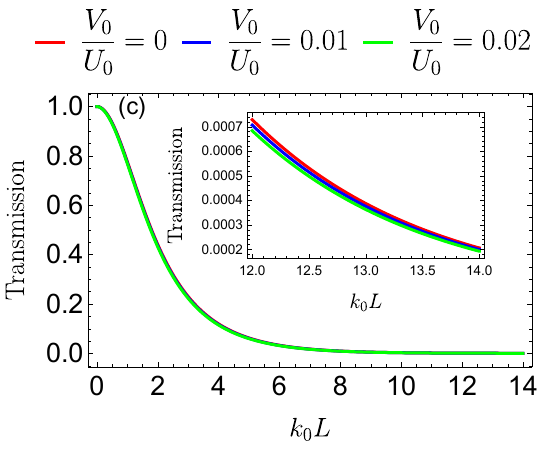}
\includegraphics[scale=0.42]{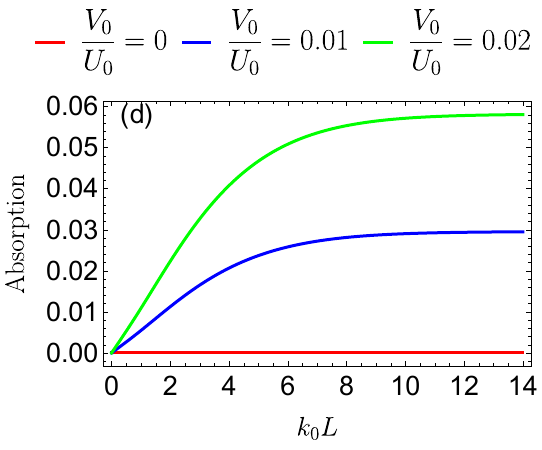}
   \caption{Transmission and absorption for a non-relativistic  particle traversing a static and noisy barrier as a function of $k_0L$. With $k_0=\sqrt{2mU_0}$. The top row corresponds to electrons with energy $\epsilon=1.1\,U_0$ that is larger than the barrier height: (a) transmission $T$ and (b) absorption $A$. The bottom row corresponds to the tunneling regime $\epsilon=0.9\,U_0$: (c) transmission $T$ and (d) absorption $A$. Curves are shown for noise strength $V_0/U_0=0$ (red), $0.01$ (blue), and $0.02$ (green).}
    \label{fig:T_A_L_Schrodinger}
\end{figure}

To visualize the analytical results and highlight the qualitative effect of noise, we now plot the transmission $T$ and absorption $A$ expressed in Eqs.~(\ref{eq:T_sch_noise}) and (\ref{eq:A_sch_noise}), respectively. We compare the case of a static barrier plus noise, $U_0\neq 0$ and $V_0\neq 0$, to the case of a purely static barrier, which corresponds to turning off the noise, $V_0=0$, and $U_0$ remains finite. We illustrate in Fig.~\ref{fig:T_A_L_Schrodinger} how these tunneling probabilities change with the barrier thickness to demonstrate the emergence of an attenuation length and the gradual disappearance of Fabry-P\'erot oscillations.

In panels (a) and (b) of Fig.~\ref{fig:T_A_L_Schrodinger}, which show the above-barrier regime $\epsilon>U_0$, in panel (a), the absence of noise ($V_0=0$), the red curve shows well-known Fabry-P\'erot oscillations that survive at any barrier thickness $L$. Amplitudes reach $1$, corresponding to total transmission. In panel (b), we find that absorption for $V_0=0$ is always zero. This observation reflects the unitary regime with probability conservation $R+T=1$.

As soon as the noise is turned on ($V_0 \neq 0$), the behavior changes qualitatively. Now, oscillations in the transmission survive only for small barrier widths $L$. Moreover, amplitudes drop exponentially with increasing barrier width. Our blue and green curves in panel (a), therefore, confirm the behavior we expect from $T(L\to\infty) \propto e^{-2 q_2 L}$, and we observe that thick barriers become effectively opaque.

This observation is easily understood from the blue and green curves in panel (b), where we see that as noise is introduced, a part of the incident flux is absorbed -the system is dissipative. Thick barriers with larger $L$ have higher absorption because particles spend more time in a lossy environment. As analytically predicted in Eq.~\eqref{eq:R_sch_noise}, we observe that the absorption saturates at a finite value less than unity.

Next, we examine the outcomes in panels (c) and (d) of Fig.~\ref{fig:T_A_L_Schrodinger}, which show the tunneling regime, $\epsilon<U_0$. For both cases, in the absence or the presence of noise, all the curves in panel (c) show that the transmission almost overlap across the full range of $k_0L$, and this transmission shows rapid, monotonic suppression with increasing barrier thickness. In particular, $T$ drops from values close to $1$ for very small $k_0L$ to values $\ll 1$ once $k_0L$ exceeds a few units ($T\lesssim 10^{-2}$ by $k_0L\sim 7$). In contrast to tunneling across a thick barrier, no Fabry-P\'erot oscillations are visible, which is consistent with the evanescent propagation mode in this regime. The common profile between the case with and without noise indicates that transmission is weakly affected, suggesting that, in this regime, the dominant suppression of transmission is already caused by the static barrier, so the wave penetrates only a short distance into the barrier. The additional damping induced by noise does not result in a quantitative reduction of $T$.

We now turn attention to panel (d), which shows the corresponding absorption. For $V_0=0$, the absorption is identically zero, which reflects the persistence of the unitary evolution and the probability conservation as seen earlier for the higher energy regime. For $V_0\neq 0$, a finite absorption develops and increases with $k_0L$ before saturating at a small value about $A\simeq 0.056$ for $\frac{V_0}{U_0}=0.01$ and $A\simeq 0.108$ for $\frac{V_0}{U_0}=0.02$. Moreover, this saturation is expected once the barrier thickness exceeds the evanescent penetration length; essentially, no additional probability weight reaches deeper into the noisy region, so increasing $L$ further does not lead to further losses. In this regime, the incoming flux is therefore predominantly reflected, with only a small fraction absorbed by the noise barrier.

\begin{figure}[ht!]
\centering 
\includegraphics[scale=0.42]{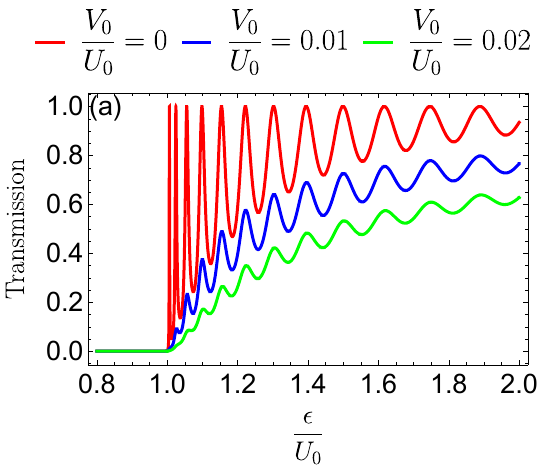}
\includegraphics[scale=0.42]{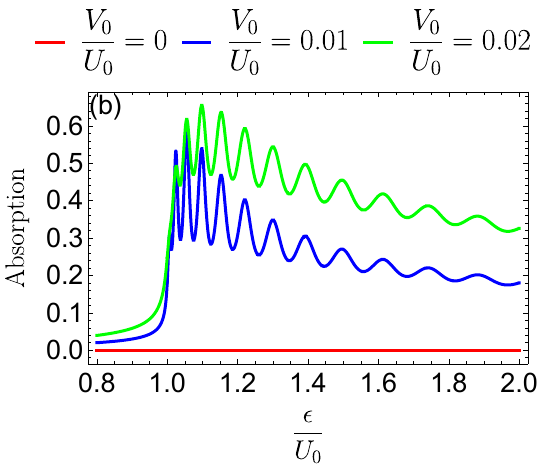}\\
\includegraphics[scale=0.42]{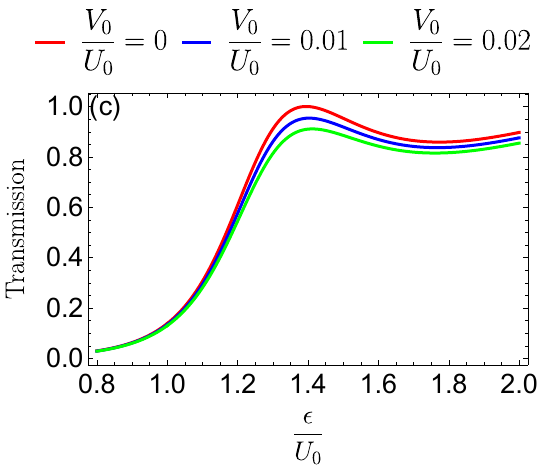}
\includegraphics[scale=0.42]{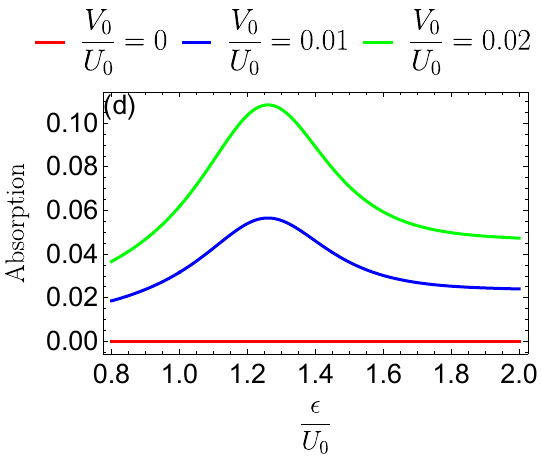}
   \caption{Transmission and absorption probabilities for a non-relativistic particle traversing a noisy barrier as functions of the dimensionless energy $\epsilon/U_0$. The top row corresponds to a thick barrier, $k_0L=40$: (a) $T$ and (b) $A$. The bottom row corresponds to a thin barrier, $k_0L=5$: (c) $T$ and (d) $A$. $V_0/U_0=0$ (red), $0.01$ (blue), and $0.02$ (green).}
    \label{fig:T_A_E_Schrodinger}
\end{figure}

We have seen that the results in Fig.~\ref{fig:T_A_L_Schrodinger} highlight how noise introduces an attenuation length as the barrier thickness increases. Now, it is also instructive to examine how the same mechanism depends on the incoming energy (keeping the barrier height and thickness fixed). Specifically in Fig.~\ref{fig:T_A_E_Schrodinger}, we study how tunneling probabilities (transmission $T$ and absorption $A$) for several values of the noise strength $V_0$ depend on dimensionless energy $\epsilon/U_0$.

Panels (a) and (b) of Fig.~\ref{fig:T_A_E_Schrodinger} show the transmission and absorption as functions of energy for a thick barrier, $k_0L=40$. First, we focus on the red curve, free of noise. In panel (a), we observe familiar Fabry-P\'erot oscillations with peaks that reach unit values. Peaks are separated by minima consistent with destructive interference inside the static barrier. In panel (b), the corresponding absorption is zero regardless of the ratio $\epsilon/U_0$, i.e., probability is conserved $R+T=1$.

Once we turn on the noise term $V_0\neq0$, we find that the blue and green curves. In panel (a), the resonant structure is strongly modified - peaks no longer reach unit values. This effect, as mentioned earlier, occurs because of the imaginary part of the wave vector $q_2>0$. We emphasize that this is because it leads to a damping effect that suppresses constructive interference. We also observe that oscillations are progressively washed out as $V_0$ increases, because noise effects dominate over barrier geometry effects.

The results in panel (b) complete this picture, where both green and blue curves show that a probability of absorption (absorption in the sense of missing coherent flux) can result from a finite noise. This result demonstrates that for energies far below the barrier height, the barrier becomes increasingly opaque. Interestingly, this means that waves are reflected before they can be absorbed, and therefore, there is no absorption - most of the incident flux is reflected at the first interface. Once the energy exceeds the barrier potential, particles penetrate deeper into the barrier, where we observe increased absorption, which reaches a maximum at an energy around the onset, $\epsilon\simeq U_0$, before decreasing. The decrease in absorption we observe at high energies is due to the increasing velocity of the incident particles. Specifically, the higher velocity means they spend less time inside the barrier and are therefore less likely to be absorbed.


Panels (c) and (d) of Fig.~\ref{fig:T_A_E_Schrodinger} show the transmission and absorption as a function of energy, but now for a much thinner barrier, $k_0L=5$. This step allows us to examine the effect of noise without the very sharp Fabry-P\'erot resonances that appear in the thick-barrier case. We see from the red curve in panel (c) (no noise, $V_0=0$) that the transmission remains small for energies below the barrier height, then rises rapidly as the energy approaches the barrier height. Interestingly, near $\epsilon\simeq U_0$, it reaches values close to $1$, and values remain close to unity for the remainder of the shown energy range. A comparison to the thick-barrier case reveals that the resonant structure is strongly broadened, and the interference features do not appear as narrow peaks but only as a smooth maximum and a weak modulation at higher energies. Once we turn on the noise ($V_0\neq 0$), this behavior changes slightly. We see in panel (c) that both the blue and green curves show that transmission is reduced compared to the non-noisy case, and that, with increasing noise strength, it reduces further. In particular, the near-unity transmission around $\epsilon\simeq U_0$ is no longer achieved. This behavior is consistent with our analytical result, where we see that the complex wavevector inside the noisy $q=q_1+i q_2$ with $q_2>0$, produces a decay and suppresses the constructive interference. 

Now, we turn our attention to panel (d). In the noiseless case ($V_0=0$, red curve), the absorption is always zero regardless of the values of energy, which is reflective of unitary time evolution and probability conservation. For finite noise ($V_0\neq 0$), a finite absorption appears. It is shown that both blue and green curves exhibit a clear maximum at energies slightly above $\epsilon\simeq U_0$. Moreover, at this range, the electrons enter the barrier efficiently and spend enough time in the noisy region, leading to the strongest absorption. For higher energies, the absorption decreases again. It approaches a weaker plateau, which is consistent with the fact that the damping wavevector $q_2(\epsilon)$ is proportional to $1/q_1$ as the real part of the wavevector $q_1$ increases. Overall, as expected, the absorption probability increases with noise strength.

\subsection{Results for graphene}

In this subsection, we apply the same density matrix approach to study noisy tunneling in graphene. Here, the impact of noise on tunneling rates, as we will see, is more striking than in the 1D Schr\"odinger case - without noise, the chirality of Dirac-Weyl fermions guarantees the Klein paradox, noise changes this. Like in the previous section, we begin our discussion with an effective stationary Lindblad Master equation in position space, which is obtained if we express Eq.~\eqref{eq:L_Dirac} in terms of a position basis.
\begin{widetext}
\begin{align}
(\epsilon-\epsilon')\,\rho(r,r') &=
\left( v_F\,\boldsymbol{\sigma}\cdot\mathbf{k}
+U(x)-i\frac{V(x)^2}{2}\right)\rho(r,r') 
-\rho(r,r')\left( v_F\,\boldsymbol{\sigma}\cdot\mathbf{k}'
+U(x')+i\frac{V(x')^2}{2}\right) \\
&+iV(x) \rho(\mathbf {r},\mathbf {r'})\,V(x') .
\label{eq:rho_graphene_position}
\end{align}
\end{widetext}
For brevity we defined $\mathbf{r}=(x,y)$ and $\mathbf{r'}=(x', y')$.

Again, along the transport direction, the static and noisy barriers $U(x)$ and $V(x)$  are piecewise constant. Therefore, in each region we may seek solutions for the density matrix that can be factored as a tensor product $\rho_{ij}(\mathbf {r},\mathbf {r'})=\rho_i(\mathbf {r})\otimes \rho_j^{\dagger}( \mathbf {r'})$. Inserting this ansatz into Eq.~\eqref{eq:rho_graphene_position} allows us to convert it into a pair of independent stationary Dirac equations for each region of Fig.~\ref{fig:nine_regions}. 

As before, the hermiticity of the density matrix imposes a strong constraint on allowed solutions, reducing the number of scattering parameters and guaranteeing the physical meaning of the derived observables, such as the expectation values of the current operator used later in the derivation of the tunneling parameters. Explicit expressions for the density matrix in the different regions of Fig.~\ref{fig:nine_regions} are relegated to Appendix \ref{app:Graphene_rho_nine_regions} for brevity.

The density matrix depends on several scattering coefficients, which are fixed by imposing continuity of the density matrix at the interfaces of the nine regions in Fig.~\ref{fig:nine_regions}. We emphasize that in the case of graphene, unlike the Schr\"odinger case, we match only the density matrix itself, without the need to match its derivatives, because graphene's Lindblad equation is first order in spatial derivatives. A detailed discussion of the matching procedure is given in Appendix~\ref{app:Graphene_matching}.

Again, to extract the expressions for tunneling probabilities, we use the probability current operator, as it is most compatible with a density-matrix formulation. That is, the expectation value of the current operator is
\begin{align}
\langle J_x \rangle= -e v_F Tr[ \rho .\sigma_x].
\end{align}
Here, $\rho$ is the density matrix for a region of interest. To obtain the incident, reflected, and transmitted currents, we insert appropriate expressions for the density matrix and find
\begin{align}
\langle J_{x}\rangle_{in} &=2e v_F \cos(\phi),\label{eq:graphene_J_in}\\
\langle J_{x}\rangle_{re} &=-2rr^{*}e v_F\cos(\phi),\label{eq:graphene_J_re}\\
\langle J_{x}\rangle_{tr} &=2 tt^{*} e v_F \cos(\phi) \label{eq:graphene_J_tr}.
\end{align}
where $\phi=\arctan\left(\frac{k_y}{k_x}\right)$ indicates the incident angle in the leads regions in Fig.~\ref{fig:nine_regions}. Following analogous steps to the Schr\"odinger case, the computed current probability in  Eqs.~(\ref{eq:graphene_J_in}),~(\ref{eq:graphene_J_re}), and ~(\ref{eq:graphene_J_tr}) enabled us to derive the transmission and reflection probabilities
\begin{align}
&T=\frac{\langle J_x \rangle_{tr}}{\langle J_x \rangle_{in}}=tt^{*}
,\qquad
R =\frac{\langle J_x \rangle_{re}}{\langle J_x \rangle_{in}}=rr^{*}.
\end{align}
Using the explicit formulas of the tunneling amplitudes $t$ and $r$, as detailed in Appendix~\ref{app:Graphene_matching}, we found  analytical expressions for the tunneling probabilities of electrons in graphene through a barrier with white noise
    \begin{align}
T &= 
\frac{A_0 e^{-2Q_2 L}}{B_0 e^{-4Q_2 L} + B_1 e^{-2Q_2 L} + B_2},
\label{eq:T_graphene_noise} \\
R &= 
\frac{C_0 \left( 1 - 2 e^{-2Q_2 L} \cos(2Q_1 L) + e^{-4 Q_2 L} \right)}
     {B_0 e^{-4Q_2 L} + B_1e^{-2Q_2 L} + B_2}.
\label{eq:R_graphene_noise}
\end{align}
Because of the non-unitary evolution induced by the noisy barrier, a part of the incident flux is absorbed:
\begin{align}
		A=1-T-R.
        \label{eq:A_graphene_noise}
	\end{align}
    Note that the Lindblad evolution preserves $\mathrm{Tr}(\rho)$
 and $A$ quantifies flux missing from the coherent plane-wave channels.\\
Here, $Q=Q_1+iQ_2$ is the complex wave vector inside the mixed static and noisy barrier, with $Q_1$ corresponding to the oscillatory propagation, and  $Q_2>0$ describes the decay induced by noise.
\begin{widetext}
\begin{align}
&Q_1 =-\frac{1}{\sqrt{2}}\left(
-k_y^2+\left(\frac{\epsilon-U_0}{ v_F}\right)^2
-\left(\frac{V_0^2}{2v_F}\right)^2
+\sqrt{
\left(
-k_y^2+\left(\frac{\epsilon-U_0}{ v_F}\right)^2
-\left(\frac{V_0^2}{2v_F}\right)^2
\right)^2
+\left(\frac{(\epsilon-U_0)V_0^2}{ v_F^2}\right)^2
}
\right)^{\frac{1}{2}},
\\
&Q_2 =\frac{(\epsilon -U_0) V_0^2}{2Q_1v_F^2},
	\end{align}
\end{widetext}
The remaining coefficients in Eqs.~(\ref{eq:T_graphene_noise}) and (\ref{eq:R_graphene_noise}) are given as \newpage
\begin{widetext}
\begin{align}
    A_0 &= 4 \cos^2(\phi) \left( \alpha_{-}^2 + 2 \alpha_{-} \alpha_{+} \cos(\theta_{-} + \theta_{+}) + \alpha_{+}^2\right),\\
	B_0 &= (\alpha_{-}^2 - 2 \alpha_{-} \cos(\theta_{-} - \phi) + 1),\\
	B_1&= 2\Big[
	\alpha_{-}^2 \cos(2(Q_1 L + \phi))
	+ 2 \alpha_{-} \sin(2Q_1 L + \phi)(\alpha_{-} \alpha_{+} \sin\theta_{+} + \sin\theta_{-}) \notag\\
	&\quad - 2 \alpha_{+} \sin(2Q_1 L - \phi)(\alpha_{-} \alpha_{+} \sin\theta_{-} + \sin\theta_{+})
	+ \alpha_{+}^2 \cos(2Q_1 L - 2\phi)
	\Big] \\
	&\quad - 2 \cos(2Q_1 L)
	(\alpha_{-}^2 \alpha_{+}^2 + 4 \alpha_{-} \alpha_{+} \sin\theta_{-} \sin\theta_{+} + 1
	(\alpha_{+}^2 - 2 \alpha_{+} \cos(\theta_{+} - \phi) + 1), \notag\\
	B_2 &= (\alpha_{-}^2 + 2 \alpha_{-} \cos(\theta_{-} + \phi) + 1)
	(\alpha_{+}^2 + 2 \alpha_{+} \cos(\theta_{+} + \phi) + 1),\\
	C_0 &= 
	\left( 
	\alpha_{-}^2 + 2 \alpha_{-} \cos(\theta_{-} + \phi) + 1 
	\right)
	\left( 
	\alpha_{+}^2 - 2 \alpha_{+} \cos(\theta_{+} - \phi) + 1 
	\right).
\end{align}
\end{widetext}
Here, quantities $\theta_\pm$ and $\alpha_\pm$ are related to the pseudospin structure (more on this below) of the forward and backward propagating modes inside the barrier and are given as
\begin{align}
\theta_{\pm} &= \arctan\!\left(\frac{k_y \pm Q_2}{Q_1}\right)
\mp \frac{1}{2}\arctan\!\left(\frac{2 Q_1 Q_2}{Q_1^2 - Q_2^2 + k_y^2}\right),\\[2mm]
\alpha_{\pm} &=
\frac{\left[Q_1^2 + (k_y \pm Q_2)^2 \right]^{1/2}}
{\left[\left(Q_1^2 - Q_2^2 +k_y^2\right)^2+ (2 Q_1 Q_2)^2 \right]^{1/4}}.
\end{align}
More precisely, $\theta_\pm$ play the role of effective propagation angles (phases) associated with the forward/backward propagating modes. Terms $\alpha_\pm$ fix the relative weight of the two pseudospin components in these modes. In the noiseless limit, when $V_0=0$, we have $Q_2=0$ and $Q_1=q_x$, where
\begin{align}
q_x=\sqrt{-k_y^2+\left(\frac{\epsilon-U_0}{v_F}\right)^2}.
\end{align}
In this case, the barrier modes reduce to the standard plane-wave solutions of a purely static barrier, so that $\theta_{+}=\theta_{-}=\arctan\left(\frac{k_y}{q_x}\right)$ and $\alpha_{+}=\alpha_{-}=1$. Accordingly, in this limit Eqs.~(\ref{eq:T_graphene_noise})--(\ref{eq:R_graphene_noise}) simplify drastically and we recover the usual unitary result with $A=0$ and $R+T=1$.  It might seem tempting to directly compute conductance by integrating over incoming momenta $k_y$ using the Landauer-B\"uttiker formula\cite{PhysRevLett.57.1761,Beenakker_1991}. However, one should keep in mind that this formula was derived in a unitary, current-conserving context and may therefore not be directly applicable to our case, which is modeled with an effective Lindblad description. Of course, one consequence is that it is not straightforward to see how interesting effects that appear directly in the conductance and are not clearly visible without angle averages integrals (like the sub-Sharvin limit observed in graphene \cite{PhysRevB.104.165413}) are modified by noise.

For graphene, if we interpret our results, we find that noise has a similar effect to that seen earlier in the Schr\"odinger case. Inside the barrier, the longitudinal wavevector is again complex, $Q = Q_1 + iQ_2$, with $Q_1$ controlling the oscillatory propagation of the Dirac spinor. Again, the imposition of the condition $Q_2 > 0$ ensures an exponential decay and defines the damping length $\ell_{\mathrm{dec}} = 1/Q_{2}$, which characterizes the length scale over which the transmitted flux is suppressed inside the noisy region.

Indeed, in the limit of thick barriers $L\to\infty$ the term $B_{2}$ in Eq.~\eqref{eq:T_graphene_noise} dominates and the transmission takes the simple asymptotic form 
\begin{equation}
T(L\to\infty) \simeq \frac{A_{0}}{B_{2}}\,e^{-2 Q_{2} L} .  
\end{equation}
Similarly, for the reflection coefficient Eq.~\eqref{eq:R_graphene_noise}, we find that exponential terms vanish in the limit $L \to \infty$, and that reflection saturates at a finite value
\begin{equation}
    R(L\to\infty) = \frac{C_{0}}{B_{2}}\leq 1.
\end{equation}

Moreover, the absorption probability $A$ (absorption in the sense of missing coherent flux) in Eq.~\eqref{eq:A_graphene_noise} acquires a finite value $A_{\infty} = 1 - R_{\infty} > 0$. Therefore, we see that in the case of finite noise $V_0\neq0$ and a thick barrier $L\to\infty$ only a finite fraction of incoming flux is reflected, and the remaining part is absorbed (for strictly $L\to\infty$, we have $T\to0$).

Of course, we also recover the well-known non-noisy results for a static barrier when $V_0=0$. Here, all the exponential terms in Eqs.~\eqref{eq:T_graphene_noise}–\eqref{eq:R_graphene_noise} become 1 because  $Q_2=0$. Therefore, there is no absorption, and we recover the typical expression for probability conservation: $R+T=1$.

A crucial point for our discussion is that this exponential suppression persists even at normal incidence, $\phi=0$. In the absence of noise, this corresponds to the channel in which fermions are perfectly transmitted through any purely static electrostatic barrier, a property related to chirality conservation and responsible for Klein tunneling. However, in our noisy-barrier problem, this protection is lifted. Therefore, to make it explicitly clear, we set $\phi=0$ ($k_y=0$). In this limit, the barrier-mode parameters simplify to $\alpha_{+}=\alpha_{-}=1$, and $\theta_{+}=\theta_{-}=0$, so that the coefficients in Eqs.~(\ref{eq:T_graphene_noise})--(\ref{eq:R_graphene_noise}) reduce to $A_0 = 16$, $B_0=0$, $B_1=0$, $B_2=16$,and $C_0$=0 . Therefore, tunneling probabilities at normal incidence take the particular form

\begin{align}
T(\phi=0) &= e^{-2Q_2 L},\label{eq:T_normal_incidence}\\
R(\phi=0) &= 0,\label{eq:R_normal_incidence}\\
A(\phi=0) &= 1-e^{-2Q_2 L}.\label{eq:A_normal_incidence}
\end{align}
From Eq.~(\ref{eq:T_normal_incidence}), we can easily conclude that any finite noise strength ($Q_2>0$) produces an attenuation length $\ell_{\mathrm{dec}}=1/Q_2$, and as $L$ increases, it turns the normally incident channel from perfectly transmitting into an exponentially suppressed one. At the same time, the missing probability is not converted into reflection but into absorption, Eq.~(\ref{eq:A_normal_incidence}), which directly identifies that noise will induce dissipation as the mechanism that circumvents the usual chirality-based protection. This result is central because it provides a route to control the channel that cannot be suppressed by static barrier modulation. In practice, this means that a temporally fluctuating barrier can act as an effective switching element, achieved by tuning the noise strength $V_0$. It allows us to continuously tune the attenuation length and reduce the transmitted current even at $\phi=0$ without changing the device geometry.

Another interesting limiting case is obtained by tuning the barrier region to its local Dirac point $\epsilon\to U_0$. This case corresponds to the pseudodiffusive regime of clean graphene discussed in \cite{PhysRevLett.96.246802,Katsnelson2006zitt}. Taking this limit (care has to be taken because it is singular) yields
\begin{equation}
\begin{gathered}
T(\phi)=\frac{1}{\cosh ^2(L \kappa)+\left(1-\frac{\Gamma^2}{\kappa^2}\right) \sinh ^2(L \kappa) \tan ^2 \phi},
\end{gathered}
\end{equation}
where we used
\begin{equation}
\begin{gathered}
\kappa^2=k_y^2+\Gamma^2, \quad \Gamma=\frac{V_0^2}{2  v_F}.
\end{gathered}
\end{equation}
as shorthand notations.\\
In the noiseless limit $V_0=0$, this reduces to the clean finite-angle evanescent result. With the additional heavily doped-lead approximation ($\phi\to0$), one recovers the standard pseudodiffusive transmission ($T(k_y)=\operatorname{sech}^2(|k_y|L)$)\cite{PhysRevLett.96.246802}.
Next,, we recall that we are dealing with an effective Lindblad system, for which the Landauer-B\"uttiker formalism is not directly justified. So while it might be tempting, we do not compute the resulting conductance. However, we can still make some observations that relate our result to previous work. For instance, in the limit of weak noise $\Gamma L\ll1$, one obtains
\begin{equation}
    T(V_0\ll1)-T(V_0=0)=c(k_y)V_0^4,
\end{equation}
where $c(k_y)<0$ is a negative function of $k_y$ (verified numerically for the full angle range $\phi$ and $k_yL$). That is, in this regime, tunneling is also similar to the noiseless case, with relatively weak suppression.

Overall, the noisy barrier behaves as a dissipative system, allowing non-unity transmission for both normal and non-normal incidence via an absorption channel that bypasses chirality conservation (absorption in the sense of missing coherent flux). 

In the following, we analyze the angular dependence of the transmission in more detail and see how noise modifies graphene's characteristic tunneling pattern.
\begin{figure}[ht!]
\centering 
\includegraphics[scale=0.5]{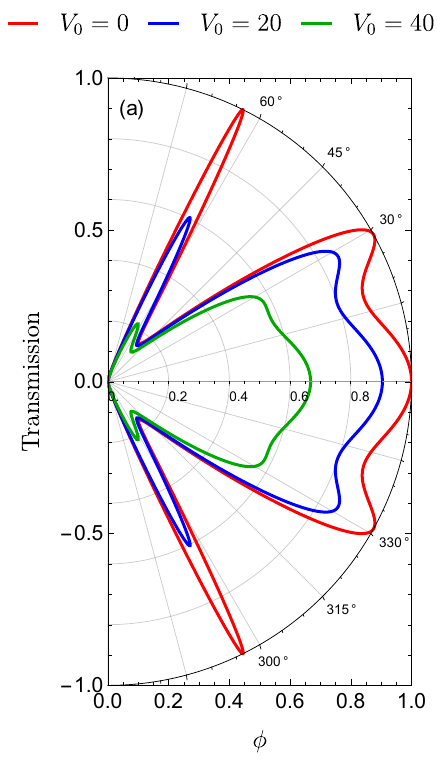}
\includegraphics[scale=0.5]{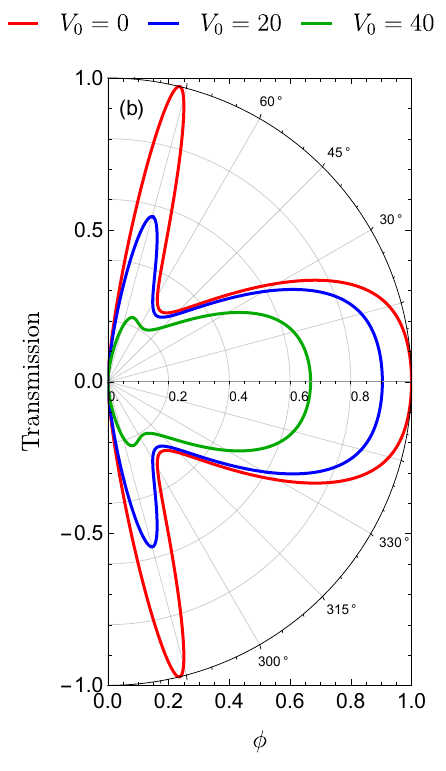}\\
\includegraphics[scale=0.5]{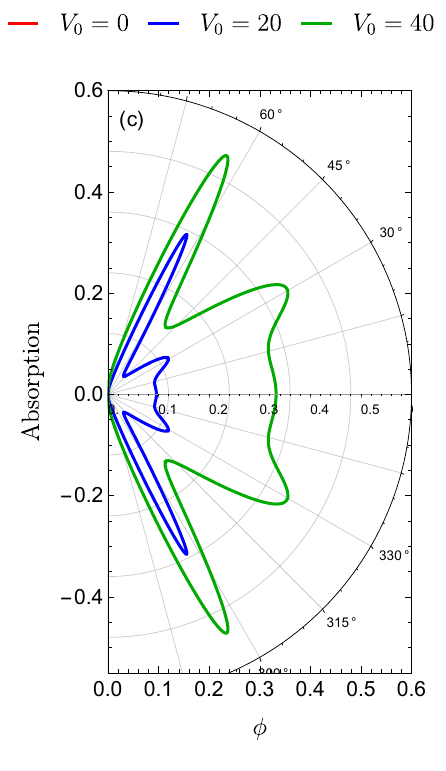}
\includegraphics[scale=0.5]{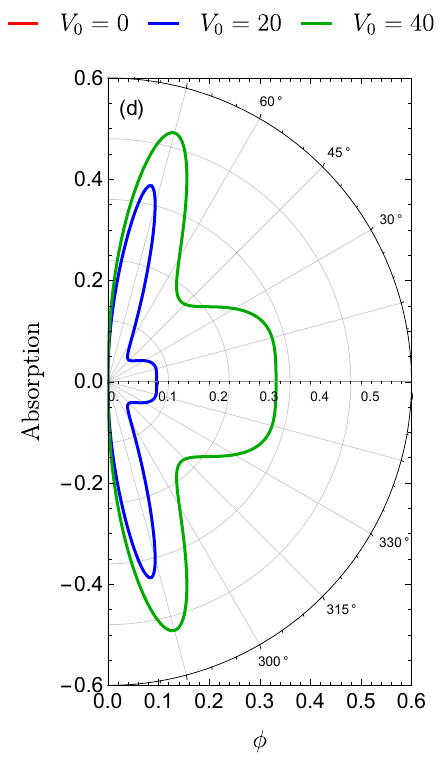}
        \label{fig:A_phi_sub}
 \caption{
    Angular dependence of transmission and absorption probabilities for tunneling through
    a noisy barrier in graphene. Panels (a,b) show the transmission $T(\phi)$ for $U_0=200$ and
    $285$~meV, and panels (c,d) show the corresponding absorption $A(\phi)$. The
    incoming energy is $\epsilon=80$~meV, the barrier width is $L=110$~nm. Curves correspond to the noise strength $V_0=0$, $20$, and
    $40$~meV.Parameters are chosen to facilitate comparison with standard Klein-tunneling setups in the literature, see Ref.~\cite{CastroNeto2009}.}
\label{fig:phi_TA}
\end{figure}
Fig.~\ref{fig:phi_TA} illustrates the effect of noise on the angular dependence of tunneling in graphene. The red curves in the top panels show that when noise is absent ($V_0=0$), at normal incidence the transmission is unity, $T(\phi=0)=1$. This expression is the famous Klein tunneling result. We observe that for a non-zero incident angle, tunneling probabilities differ from unity. Dips and peaks in the probability distribution arise from resonance conditions within the static barrier. Our numerical results in this case agree exactly with those reported in Ref.~\cite{CastroNeto2009}.

As soon as noise is switched on, $V_0>0$ (see the blue and green curves in the top panels), we observe that the transmission is reduced at all incidence angles, and the sharp resonant features of the static barrier are washed out. Most strikingly, the perfect transmission at normal incidence, $T(\phi=0)$, is suppressed, dropping below unity and decreasing further as $V_0$ increases. Overall, our results demonstrate that the noisy barrier leads to an exponential suppression of the transmission probability for both normal and oblique incidence. This observation is consistent with what we have seen earlier in the analytical expressions of Eqs~(\ref{eq:T_graphene_noise}),~(\ref{eq:T_normal_incidence}), where the transmission acquires an exponential decaying factor $T \propto e^{-2 Q_2 L}$ with $Q_2>0$. 

To complete the picture, we now turn our attention to the lower panels (c) and (d) in Fig.~\ref{fig:phi_TA}, which show how absorption depends on the angle of incidence. In the noiseless case, where $V_0=0$ (red curves), the absorption is zero for all angles (it is therefore invisible in the plot). This observation is consistent with tunneling through the static barrier, where the probability current is conserved and $R+T=1$, so there is no loss channel. Both blue and green curves show that in the presence of noise $V_0>0$, the absorption probability becomes finite as a result of dissipation caused by the noisy barrier, and the absorption profile shows a pronounced angular dependence. In particular, it is non-zero at normal incidence, which demonstrates that the noisy barrier opens a loss channel even when reflection remains suppressed, as we have seen in Eq.~(\ref{eq:R_normal_incidence}). The absorption typically reaches higher values at certain angles, where a sizable fraction of the incoming flux penetrates the barrier while the effective path length through the lossy region increases. For larger incidence angles approaching grazing incidence, penetration into the barrier is strongly reduced due to enhanced reflection at the first interface, and the absorption decreases accordingly. In the main text, we discuss only the tunneling regime $E<U_0$, since results for the over-the-barrier regime $E>U_0$ do not differ significantly. Relevant plots are in appendix \ref{app:tunneling}.

\begin{figure}[ht!]
\centering 
\includegraphics[scale=0.45]{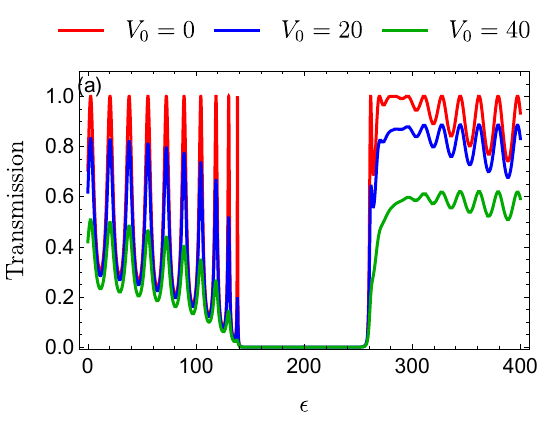}
\includegraphics[scale=0.45]{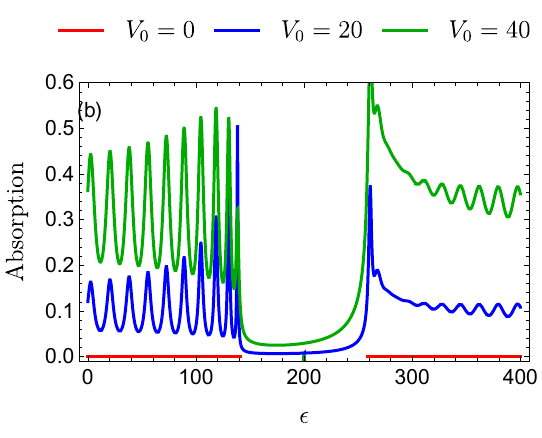}

 \caption{Transmission (a) and absorption (b) probabilities for electrons in graphene tunneling through a noisy barrier as a function of incoming energy for an incident angle $\phi=45^\circ$, barrier width  $L= 110~nm$, and static barrier height is  ~$U_0 =200~meV$. The three curves correspond to noise strength $V_0 = 0$ (red), $20$ (blue), and  $40$~meV(green).
    }
\label{fig:T_A_E_Graphene}
\end{figure}

So far, we have focused on the angular dependence, which makes clear that noise opens a loss channel and suppresses the Klein-tunneling peak - even at $\phi=0$. It is also instructive to examine how this mechanism depends on the incoming energy at a fixed geometry and a fixed incidence angle. Therefore, in Fig.~\ref{fig:T_A_E_Graphene} we plot the transmission and absorption as functions of the incoming energy $\epsilon$, while keeping $U_0$ at 200 meV and $L$ constant at $110~nm$, and $\phi=45^\circ$, by varying the noise strength $V_0$.

The results in panel (a) of Fig.~\ref{fig:T_A_E_Graphene} demonstrate the energy dependence of tunneling through noisy barriers in graphene. When $V_0=0$ (red curve), the transmission in panel~(a) shows Fabry-P\'erot oscillations with peaks reaching 1. This effect arises from resonances within the static barrier. We also observe a parameter region around $200$ meV with zero transmission, corresponding to a regime in which the conserved transverse momentum required by the incident angle exceeds the kinetic energy available inside the barrier. The corresponding absorption is illustrated in panel ~(b) and zero  for energies where entry into the barrier region is possible. Our result confirms the unitary evolution and the conservation of probabilities $R+T=1$.

Once noise is switched on with $V_0>0$ (see the blue and green curves in panel (a)), this behavior changes dramatically. Peaks shrink as $V_0$ increases, and resonances never reach values even close to unity. This observation is consistent with Eq.~(\ref{eq:T_graphene_noise}), where the imaginary part of the wavevector $Q_2>0$ caused an overall damping factor $e^{-2Q_2L}$ for the Fabry-P\'erot structure.  Interestingly, we also observe that, even in the absence of tunneling, there is nonzero absorption, indicating absorption at the barrier surface.

In summary, as our central result, we observe that tunneling through a temporally noisy barrier opens an effective loss channel and introduces a finite decay length inside the barrier. This observation also holds, most importantly at normal incidence, where a purely static barrier yields perfect Klein transmission. Here, the introduction of Gaussian white noise at the barrier exponentially suppresses the tunneling probability $T \sim e^{-2Q_2L}$, where $Q_2>0$. That is, it gives rise to an attenuation length $\ell_{\mathrm{dec}}=1/Q_2$, which can be tuned by noise strength $V_0$ and other barrier parameters. Operationally, this means that by tuning the noise strength $V_0$, one can make a given barrier effectively opaque. It allows us to continuously reduce the transmitted current without any need to modify the geometry of the static barrier. In a device setting, such temporal fluctuations can arise from a deliberately modulated gate voltage or from controlled coupling of the graphene channel to a dissipative environment. Overall, the key point is that noise provides an efficient control knob that could help reduce the usual limitation imposed by Klein tunneling and enhance the switching capability of a graphene-based device.

\section{Usefulness of the approach beyond control of electron transport}
\label{sec:App}

\begin{figure}[htbp!]
    \centering
\includegraphics[width=\columnwidth]{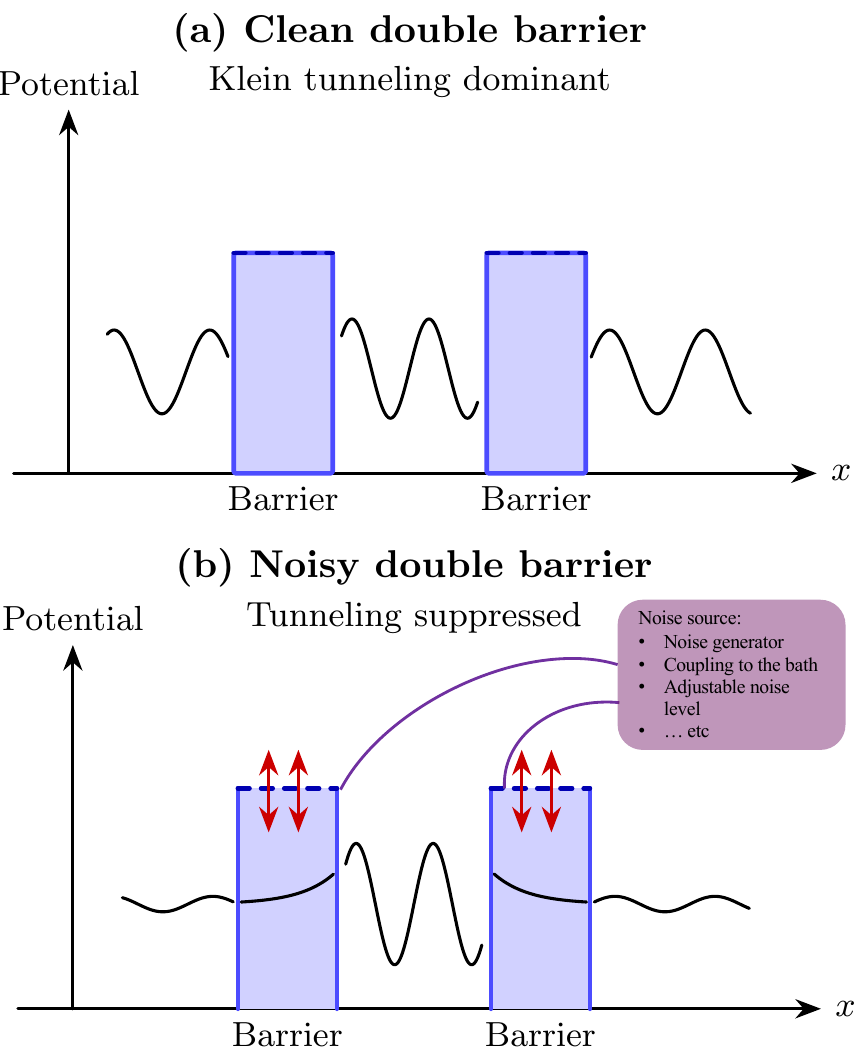}
    \caption { Illustration of Quantum confinement in two cases:(a) Clean double barrier where Klein tunneling dominates, and the particle cannot efficiently be confined. (b) A noisy double barrier, based on our calculations, could suppress Klein tunneling and confine the particle.  }
    \label{fig:noisy_confinement}
\end{figure}

Beyond the switch capability in transport, our results could also be informative for recently emerging technologies. For instance, spin qubits are typically realized by confining a single electron in a potential well, or quantum dot, formed in a semiconductor material. Such confinement can be achieved by applying gate voltages to create tunable potential barriers that localize the electron and allow precise control of its quantum state \cite{Bukarad2023}. In this context, a monolayer graphene is advantageous for quantum dot realization due to its weak spin–orbit coupling and weak hyperfine interaction, both of which contribute to large spin coherence times and robust quantum control, although its gapless spectrum requires additional strategies to achieve efficient confinement \cite{Trauzettel2007}. Our results could complement such techniques.
Specifically, Klein tunneling strongly hinders the electrostatic confinement of electrons in a potential well, as illustrated in Fig.~\ref{fig:noisy_confinement}, which limits graphene’s capability in such applications \cite{Trauzettel2007}. 
As shown in previous sections, our results suggest that a noisy or dissipative potential barrier could reduce leakage from Klein tunneling and thereby help stabilize electron confinement in multi-barrier graphene structures. A quantitative statement about dot-bound states would require an explicit double-barrier computation. Nevertheless, noise must be introduced in a controlled manner, as it may also reduce the coherence length and lifetime of the quantum state as shown in Fig~\ref{fig:noisy_confinement}. This tradeoff is closely related to the standard dissipative two-state problem, where environmental coupling can suppress coherent tunneling dynamics and induce decoherence \cite{RevModPhys.59.1,doi:10.1142/8334}. This indicates that the noise strength should be regarded as an effective barrier parameter, on equal footing with the barrier height and width. Such an approach may enable the design of optimized graphene-based spin qubits and provide a consistent theoretical framework grounded in the methods developed in this work. It may also help in better understanding the impact of noise on system parameters.
\\ 
We emphasize that our results are not intended to constitute a proposal for a complete qubit architecture. Rather, they suggest that noise-induced suppression of Klein tunneling could provide an additional mechanism that may be incorporated into analytical models of graphene quantum-dot confinement \cite{Trauzettel2007} and complement existing approaches discussed in the literature.\\

Recently, many experiments have focused on measuring and suppressing noise to increase qubit coherence time \cite{Bukarad2023,BarGill2012,Soare2014}. This suggests that the effect can be experimentally investigated in engineered monolayer graphene-based devices. One method employed in semiconductor quantum dots to suppress noise is bias cooling, in which a voltage is applied to the gate while the system cools down \cite{Ferrero2024}. These studies suggest that charge-noise environments in gate-defined quantum-dot devices can be engineered to some extent, although implementing controlled broadband noise in graphene barriers would require a dedicated device design.

\section{Conclusion}
\label{sec:conclusion}

In this work, we study tunneling across a noisy barrier whose height fluctuates in time according to Gaussian white noise. By averaging over noise realizations within the Lindblad formalism, we recast the problem as an effective stationary scattering problem. This reformulation removes the need for computationally expensive numerical simulations and allows us to derive analytical expressions for the tunneling probabilities.

First, we studied tunneling for a non-relativistic particle as described by the Schr\"odinger equation. We found that the noisy barrier induces a complex wave vector, leading to exponential decay in transmission. For non-zero noise, the sum of reflection and transmission coefficients $T+R$ saturates at a value $\leq 1$, and the missing part of the incoming flux is absorbed. Therefore, the Schr\"odinger case provides a useful minimal example of how noise can induce dissipation in a tunneling scenario.

Second, we applied the density matrix formalism to investigate the impact of noise on tunneling in graphene. The results show that averaging over noise realizations again leads to damping within the barrier region, resulting in an exponentially decaying transmission coefficient even at normal incidence. In other words, we were able to suppress Klein tunneling. Our findings suggest that the strongest absorption, interpreted as the loss of coherent flux near the barrier, can serve as a control knob for electron transport in graphene.

From a methodological perspective, the main contribution of this work is the combination of an exact mapping from Gaussian white-noise dynamics to a stationary Lindblad equation with a fully analytical solution of the resulting higher-dimensional scattering problem. This approach provides closed-form expressions that explicitly reveal how stochastic temporal fluctuations suppress Klein tunneling through dissipative attenuation.

The present methodology opens the door to several future research directions. For instance, one could study other types of noise, such as non-Gaussian noise, or go beyond white noise and consider colored noise. In the colored-noise case, even after averaging over noise realizations, the dynamics remain explicitly time-dependent (although typically smoother than for white noise). Moreover, the price to pay is that the effective master equation is no longer local in time: memory kernels appear, and the evolution becomes non-Markovian. Certainly, it will also be interesting to extend the present approach to multi-barrier structures as in our quantum dot example. A study of other more exotic semimetals, and inhomogeneous two-dimensional setups might also be interesting. 
Our main result, with potential technological implications, is that noisy or dissipative barriers can serve as an additional control mechanism for graphene-based electronic devices. By tuning the noise strength or the coupling to the surrounding environment, one can achieve enhanced control over electron transmission and confinement. In particular, Klein tunneling can be efficiently suppressed under such conditions. Additional work is needed to better identify realistic noise sources and explore concrete device geometries in which this mechanism could be experimentally realized and tested. 

\section{Acknowledgments}
  A. Al Luhaibi at KFUPM would like to acknowledge the support received under the University Funded Grant EC241024. M.V. gratefully acknowledges the support provided by the Deanship of Research Oversight and Coordination (DROC) and the Interdisciplinery Reasearch Center(IRC) for Advanced Quantum Computing (AQC) at King Fahd University of Petroleum \& Minerals (KFUPM) for funding his contribution to this work through research grant No. INQC2607. K. Azaidaoui acknowledges the support provided by CNRST in the framework of the program ``PhD-Associate Scholarship -- PASS''.

\bibliographystyle{unsrt}
\bibliography{literature}
\appendix
\label{appendix}
\section{Schr\"odinger case: detailed expressions}
\label{app:sch_full}
\subsection{Density matrix in the nine regions}
\label{app:sch_rho_nine_regions}
The stationary Lindblad equation in the position representation has constant coefficients within each of the nine regions $(i,j)$ of the $(x,x')$ plane shown in Fig.~\ref{fig:nine_regions}, because $U(x)$ and $V(x)$ are piecewise constant. As a consequence, within each region, the stationary solution can be written as a linear combination of plane-wave factors in $x$ and in $x'$. We parametrize the solution using standard scattering amplitudes in the leads. We set the incident amplitude to unity and denote the reflection and transmission amplitudes by $r$ and $t$. Additional forward and backward amplitudes are required in regions where one coordinate lies inside the noisy barrier. As explained, the hermiticity, $\rho_{ij}(x,x')=\rho_{ji}^\dagger(x',x)$, relates different regions and reduces the number of independent coefficients. With these conventions, a convenient representation of the density matrix in the nine regions in  Fig.~\ref{fig:nine_regions} is given below.
\begin{widetext}
	\begin{align}
		&\rho_{11}(x,x^{\prime}) =(e^{ikx}+re^{-ikx})(e^{-ikx^{\prime}}+r^{*}e^{ikx^{\prime}})\\
        &\rho_{21}(x,x^{\prime})=(ae^{iqx}+be^{-iqx})(e^{-ikx^{\prime}}+r^{*}e^{ikx^{\prime}})\\
        &\rho_{31}(x,x^{\prime})=te^{ikx}(e^{-ikx^{\prime}}+r^{*}e^{ikx^{\prime}})\\
        &\rho_{12}(x, x^{\prime}) = (e^{ikx} + r e^{-ikx})(a^{*}e^{-iq^*x^{\prime}}+b^{*}e^{iq^*x^{\prime}})\\
        &\rho_{22}(x,x^{\prime}) =(ce^{iq_0x}+de^{-iq_0x})(c^{*}e^{-iq_0x^{\prime}}+d^{*}e^{iq_0x^{\prime}})\\
        &\rho_{32}(x,x^{\prime})=te^{ikx}(a^{*}e^{-iq^*x^{\prime}}+b^{*}e^{iq^*x^{\prime}})\\
        &\rho_{13}(x, x^{\prime}) = (e^{ikx} + r e^{-ikx}) t^{*} e^{-ikx^{\prime}}\\
		&\rho_{23}(x, x^{\prime}) = (ae^{iqx}+be^{-iqx}) t^{*} e^{-ikx^{\prime}}\\
		&\rho_{33}(x,x^{\prime})=te^{ikx}t^{*}e^{-ikx^{\prime}}
	\end{align} 
    \end{widetext}
Here, $r$ and $t$ are the reflection and transmission amplitudes in the leads. Moreover, $a$ and $b$ describe regions where one coordinate is inside the noisy barrier, and the other is in a lead. In contrast, $c$ and $d$ describe the diagonal region $(2,2)$ where both coordinates lie inside the barrier. In that diagonal region, the Lindblad terms cancel, and propagation is governed solely by the static barrier.
Our expressions made use of the following wavevectors. In the leads we have
\begin{align}
k=\sqrt{2m\epsilon}
\end{align}
and in the diagonal region $(2,2)$
\begin{align}
q_0=\sqrt{2m(\epsilon-U_0)}.
\end{align}
In regions where one coordinate lies inside the noisy barrier, the longitudinal wavevector becomes complex, $q=q_1+i q_2$, with
\begin{align}
q_1&=\left[m(\epsilon-U_0)
+\sqrt{\left(m(\epsilon-U_0)\right)^2
+\left(\frac{mV_0^2}{2}\right)^2}\right]^{1/2},\\
q_2&=\frac{mV_0^2}{2q_1}.
\end{align}
\subsection{Matching conditions, scattering amplitudes, and probabilities}
\label{app:sch_matching}
In the Schr\"odinger case, the stationary Lindblad equation in the position representation contains second-order spatial derivatives (in $x$ and $x'$). Therefore, at interfaces where $U(x)$ and $V(x)$ change discontinuously, physical solutions require continuity of the density matrix and its first spatial derivative with respect to the transport coordinate. Imposing these conditions on the two interfaces $x=0$ and $x=L$ (and analogously for $x'$) yields matching relations at the diagonal points $(x,x')=(0,0)$ and $(L, L)$:
\begin{align}
\rho_{11}(0,0) &= \rho_{21}(0,0)=\rho_{12}(0,0), \\
\rho_{23}(L,L) &= \rho_{32}(L,L)=\rho_{33}(L,L), \\
\partial_x \rho_{11}(0,0) &= \partial_x \rho_{21}(0,0)=\partial_x \rho_{12}(0,0), \\
\partial_x \rho_{23}(L,L) &= \partial_x \rho_{32}(L,L)=\partial_x \rho_{33}(L,L).
\end{align}
The corresponding conditions involving $\partial_{x'}$ follow analogously and are not independent because of hermiticity.
Inserting explicit expressions for $\rho_{ij}$ from Appendix~\ref{app:sch_rho_nine_regions} into the matching conditions yields a closed linear system for the scattering amplitudes $r,t$ in the leads and the auxiliary coefficients $a,b$ in the regions where only one coordinate lies inside the noisy barrier:
\begin{align}
1+r &= a+b, \label{eq:sch_lin_1}\\
k(1-r) &= q(a-b), \label{eq:sch_lin_2}\\
ae^{iqL}+be^{-iqL} &= t\,e^{ikL}, \label{eq:sch_lin_3}\\
q\!\left(ae^{iqL}-be^{-iqL}\right) &= k\,t\,e^{ikL}. \label{eq:sch_lin_4}
\end{align}
Solving Eqs.~(\ref{eq:sch_lin_1})--(\ref{eq:sch_lin_4}) gives the tunneling
amplitudes
\begin{align}
t &=\frac{2 i k \,q\, e^{-i k L}}
{\left(k^2+q^2\right)\sin(qL)+2 i k q \cos(qL)},
\label{eq:sch_t_app}\\
r &=\frac{\left(k^2-q^2\right)\sin(qL)}
{\left(k^2+q^2\right)\sin(qL)+2 i k q \cos(qL)}.
\label{eq:sch_r_app}
\end{align}
For completeness, the corresponding coefficients inside the barrier regions are
\begin{align}
a &=-\frac{2k(k+q)}{(-k+q)^2 e^{2 i qL}-(k+q)^2}, \label{eq:sch_a_app}\\
b &=\frac{2k(k-q)}{(k-q)^2-e^{-2 i qL}(k+q)^2}.
\label{eq:sch_b_app}
\end{align}

Using the probability-current definition in the main text, the transmission and reflection probabilities take the form
\begin{align}
T &= \frac{4k^2(q_1^2+q_2^2)}{D_0 + D_1 e^{2q_2L} + D_2 e^{-2q_2L}},
\label{eq:sch_T_app}\\
R &= \frac{N_0\left[\cosh(2q_2L)-\cos(2q_1L)\right]}
{D_0 + D_1 e^{2q_2L} + D_2 e^{-2q_2L}}.
\label{eq:sch_R_app}
\end{align}
The coefficients in Eqs.~(\ref{eq:sch_T_app}) and (\ref{eq:sch_R_app}) are
\begin{widetext}
\begin{align}
D_0 &=
-2k q_2\left(k^2-q_1^2-q_2^2\right)\sin(2Lq_1)
-\frac12\!\left[k^4-2k^2\left(q_1^2+3q_2^2\right)+\left(q_1^2+q_2^2\right)^2\right]\cos(2Lq_1),\\
D_{1,2} &= \frac14\left[(k\pm q_1)^2+q_2^2\right]^2,\\
N_0 &= \frac12\left[k^4+2k^2\left(q_2^2-q_1^2\right)+\left(q_1^2+q_2^2\right)^2\right].
\end{align}
\end{widetext}
\section{Graphene case: detailed expressions}
\label{app:Graphene_full}
\subsection{Density matrix in the nine regions}
\label{app:Graphene_rho_nine_regions}
For graphene, the density matrix elements $\rho_{ij}(\mathbf r,\mathbf r')$ are $2\times2$ matrices in pseudospin space. We use the nine-region partition of the $(x,x')$ plane shown in Fig.~\ref{fig:nine_regions}. In each region, the stationary Lindblad equation has constant coefficients. We therefore write the solution in separable form, $\rho_{ij}(\mathbf r,\mathbf r')=\rho_i(\mathbf r)\otimes\rho_j^\dagger(\mathbf r')$, where $\rho_i$ are two-component spinors in region $i$. We set the incident amplitude to one. We parametrize the lead solutions by the reflection and transmission amplitudes $r$ and $t$, and we use additional coefficients $a$ and $b$ (mixed static+noisy barrier) and $c$ and $d$ (purely static barrier in the diagonal region $i=j=2$). Hermiticity, $\rho_{ij}(\mathbf r,\mathbf r')=\rho_{ji}^\dagger(\mathbf r',\mathbf r)$, ensures that the same set of amplitudes appears consistently across the nine regions. The resulting density matrices in each region are given as
\begin{widetext}
\begin{align}
		\rho_{11}
        =&\begin{pmatrix}
			e^{ik_xx}+re^{-ik_xx}\\
			e^{ik_xx+i\phi}-re^{-ik_xx-i\phi}
		\end{pmatrix}e^{ik_yy}\otimes\begin{pmatrix}
			e^{-ik_xx^{\prime}}+r^{*}e^{ik_xx^{\prime}},&
			e^{-ik_xx^{\prime}-i\phi}-r^{*}e^{ik_x x^{\prime}+i\phi}
		\end{pmatrix}e^{-ik_yy^{\prime}},
	\\
			\rho_{21}=&\begin{pmatrix}
				ae^{iQx} + be^{-iQx} \\
				\alpha_{+} ae^{iQx + i\theta_{+}} - \alpha_{-} be^{-iQx - i\theta_{-}}
			\end{pmatrix}
			e^{ik_y y}\otimes\begin{pmatrix}
				e^{-ik_xx^{\prime}}+r^{*}e^{ik_xx^{\prime}},&
				e^{-ik_xx^{\prime}-i\phi}-r^{*}e^{ik_x x^{\prime}+i\phi}
			\end{pmatrix}e^{-ik_yy^{\prime}},
	\\
		\rho_{31}=&\begin{pmatrix}
			te^{ik_xx}\\
			te^{ik_xx+i\phi}
		\end{pmatrix}e^{ik_yy}\otimes\begin{pmatrix}
			e^{-ik_xx^{\prime}}+r^{*}e^{ik_xx^{\prime}},&
			e^{-ik_xx^{\prime}-i\phi}-r^{*}e^{ik_x x^{\prime}+i\phi}
		\end{pmatrix}e^{-ik_yy^{\prime}},	
	\\
			\rho_{12} = &
			\begin{pmatrix}
				e^{ik_xx} + r e^{-ik_xx} \\
				e^{ik_xx+i\phi} - r e^{-ik_xx - i\phi}
			\end{pmatrix}e^{ik_yy} 
			\otimes 
			\begin{pmatrix}
				a^{*}e^{-iQ^*x^{\prime}} + b^{*}e^{iQ^*x^{\prime}},&
				\alpha_{+} a^{*}e^{-iQ^*x^{\prime} - i\theta_{+}} - \alpha_{-} b^{*}e^{iQ^*x^{\prime} + i\theta_{-}}
			\end{pmatrix}
			e^{-ik_y y^{\prime}}, 
	\\
		\rho_{22}=&\begin{pmatrix}
			ce^{iq_xx}+de^{-iq_xx}\\
			ce^{iq_xx+i\theta}-de^{-iq_xx-i\theta}
		\end{pmatrix}e^{ik_yy}\\
        &\otimes\begin{pmatrix}
			c^{*}e^{-iq_xx^{\prime}}+d^{*}e^{iq_xx^{\prime}},&
			c^{*}e^{-iq_xx^{\prime}-i\theta}-d^{*}e^{iq_xx^{\prime}+i\theta}
		\end{pmatrix}e^{-ik_yy^{\prime}},
	\\
			\rho_{32}=&\begin{pmatrix}
				te^{ik_xx}\\
				te^{ik_xx+i\phi}
			\end{pmatrix}e^{ik_yy}\\
            & \otimes
            \begin{pmatrix}
				a^{*}e^{-iQ^*x^{\prime}} + b^{*}e^{iQ^*x^{\prime}},&
				\alpha_{+} a^{*}e^{-iQ^*x^{\prime} - i\theta_{+}} - \alpha_{-} b^{*}e^{iQ^*x^{\prime} + i\theta_{-}}
			\end{pmatrix}
			e^{-ik_y y^{\prime}},
	\\
		\rho_{13} =&\begin{pmatrix}
			e^{ik_xx}+re^{-ik_xx}\\
			e^{ik_xx+i\phi}-re^{-ik_xx-i\phi}
		\end{pmatrix}e^{ik_yy}\otimes\begin{pmatrix}
			t^{*}e^{-ik_xx^{\prime}},&
			t^{*}e^{-ik_xx^{\prime}-i\phi}
		\end{pmatrix}e^{-ik_yy^{\prime}},
	\\
    \rho_{23}=&\begin{pmatrix}
			ae^{iQx} + be^{-iQx} \\
			\alpha_{+} ae^{iQx + i\theta_{+}} - \alpha_{-} be^{-iQx - i\theta_{-}}
		\end{pmatrix}
		e^{ik_y y}\otimes\begin{pmatrix}
			t^{*}e^{-ik_xx^{\prime}}&
			t^{*}e^{-ik_x x^{\prime}-i\phi}
		\end{pmatrix}e^{-ik_yy^{\prime}},
	\\
    \rho_{33}=&\begin{pmatrix}
			te^{ik_xx}\\
			te^{ik_xx+i\phi}
		\end{pmatrix}e^{ik_yy}\otimes\begin{pmatrix}
			t^{*}e^{-ik_xx^{\prime}},&
			t^{*}e^{-ik_xx^{\prime}-i\phi}
		\end{pmatrix}e^{-ik_yy^{\prime}}.
	\end{align}.
\end{widetext}
Here, $k_y$ is conserved. In the leads ($U=V=0$) we use
\begin{align}
k_x=\sqrt{\left(\frac{\epsilon}{ v_F}\right)^2-k_y^2},
\qquad
\phi=\arctan\!\left(\frac{k_y}{k_x}\right).
\end{align}
In the diagonal region $i=j=2$, both coordinates lie inside the barrier, the
Lindblad terms cancel, and the static barrier governs the propagation
$U_0$. The corresponding longitudinal wavevector is
\begin{align}
q_x=\sqrt{\left(\frac{\epsilon-U_0}{v_F}\right)^2-k_y^2},
\qquad
\theta=\arctan\!\left(\frac{k_y}{q_x}\right).
\end{align}

In the mixed static and noisy barrier, the longitudinal wavevector becomes
complex, $Q=Q_1+iQ_2$, with $Q_2>0$ setting the attenuation:
\begin{widetext}
\begin{align}
Q_1 &=-\frac{1}{\sqrt{2}}\left(
-k_y^2+\left(\frac{\epsilon-U_0}{ v_F}\right)^2
-\left(\frac{V_0^2}{2v_F}\right)^2
+\sqrt{
\left(
-k_y^2+\left(\frac{\epsilon-U_0}{ v_F}\right)^2
-\left(\frac{V_0^2}{2v_F}\right)^2
\right)^2
+\left(\frac{(\epsilon-U_0)V_0^2}{ v_F^2}\right)^2
}
\right)^{\frac{1}{2}},
\\
Q_2 &=\frac{(\epsilon-U_0) V_0^2}{2Q_1v_F^2}.
\end{align}
\end{widetext}

Phases $\theta_\pm$ and weights $\alpha_\pm$ encode the pseudospin structure
of the forward and backward modes inside the mixed static and noisy barrier:
\begin{align}
\theta_{\pm} &= \arctan\!\left(\frac{k_y \pm Q_2}{Q_1}\right)
\mp \frac{1}{2}\arctan\!\left(\frac{2 Q_1 Q_2}{Q_1^2 - Q_2^2 + k_y^2}\right),
\\
\alpha_{\pm} &=
\frac{\left[Q_1^2 + (k_y \pm Q_2)^2 \right]^{1/2}}
{\left[\left(Q_1^2 - Q_2^2 +k_y^2\right)^2+ (2 Q_1 Q_2)^2 \right]^{1/4}}.
\end{align}

\subsection{Matching conditions and scattering amplitudes}
\label{app:Graphene_matching}

We determine the scattering amplitudes by imposing continuity of the density
matrix at the interfaces of the nine-region construction in Fig.~\ref{fig:nine_regions}.
Because the graphene Lindblad equation is first order in spatial derivatives, we
match only the density matrix itself $\rho$ itself. We therefore impose the boundary conditions at
$(x,x')=(0,0)$ and $(x,x')=(L,L)$,
\begin{align}
\rho_{11}(0,0) =\rho_{21}(0,0)=\rho_{12}(0,0), \\
\rho_{23}(L,L) = \rho_{32}(L,L)=\rho_{33}(L,L).
\end{align}
Inserting the expressions from Appendix~\ref{app:Graphene_rho_nine_regions} yields a linear system for $r,t,a,b$:
\begin{align}
a+b &= 1+r, \\
\alpha_{+}\,a\,e^{i\theta_{+}}-\alpha_{-}\,b\,e^{-i\theta_{-}} &= e^{i\phi}-r\,e^{-i\phi}, \\
a\,e^{iQL}+b\,e^{-iQL} &= t\,e^{i k_{x} L}, \\
\alpha_{+}\,a\,e^{iQL+i\theta_{+}}-\alpha_{-}\,b\,e^{-iQL-i\theta_{-}} &= t\,e^{i k_{x} L+i\phi}.
\end{align}
Solving this system gives the lead amplitudes $t$ and $r$ and the internal amplitudes
$a$ and $b$:
\begin{widetext}
\begin{align}
t&=\frac{\left(1+e^{2 i \phi }\right) \left(\alpha_{-}+\alpha_{+} e^{i (\theta_{-}+\theta_{+})}\right) e^{-i(k_x-Q)L}}{\alpha_{-} + e^{i (\theta_{-}+\phi )}+\alpha_{+} e^{i \theta_{+}} \left(\alpha_{-} e^{i \phi }+e^{i (\theta_{-}+2 \phi )}-\alpha_{-} e^{i (\phi +2 L Q)}+e^{i (\theta_{-}+2 L Q)}\right)+\alpha_{-} e^{2 i (\phi +L Q)}-e^{i (\theta_{-}+2 L Q+\phi )}},
\\
r&=-\frac{e^{i \phi } \left(\alpha_-+e^{i (\theta_{-}+\phi )}\right) \left(e^{i \phi }-\alpha_{+} e^{i \theta_{+}}\right) \left(-1+e^{2i Q L}\right)}{\alpha_-+e^{i (\theta_{-}+\phi )}+\alpha_{+} e^{i \theta_{+}} \left(\alpha_- e^{i \phi }+e^{i (\theta_{-}+2 \phi )}-\alpha_- e^{i (\phi +2 L Q)}+e^{i (\theta_{-}+2 L Q)}\right)+\alpha_- e^{2 i (\phi +L Q)}-e^{i (\theta_{-}+2 L Q+\phi )}},
\\
a&=\frac{\left(1+e^{2 i \phi }\right) \left(\alpha_{-}+e^{i (\theta_{-}+\phi )}\right)}{\alpha_{-}+e^{i (\theta_{-}+\phi )}+\alpha_{+} e^{i \theta_{+}} \left(\alpha_{-} e^{i \phi }+e^{i (\theta_{-}+2 \phi )}-\alpha_{-} e^{i (\phi +2 L Q)}+e^{i (\theta_{-}+2 L Q)}\right)+\alpha_{-} e^{2 i (\phi +L Q)}-e^{i (\theta_{-}+2 L Q+\phi )}},
\\
b&=
\frac{e^{i(2LQ+\theta_-)}\,(1+e^{2i\phi})\,(e^{i\phi}-e^{i\theta_+}\alpha_+)}
{\Big(e^{i(2LQ+\theta_-)}+e^{i(\theta_-+2\phi)}+e^{i\phi}\alpha_--e^{i(2LQ+\phi)}\alpha_-\Big)\,(e^{i\phi}-e^{i\theta_+}\alpha_+)\;-\;(1+e^{2i\phi})\Big(e^{i(\theta_-+\phi)}+\alpha_-\Big)}.
\end{align}
\end{widetext}
The tunneling probabilities follow from the current ratios in the main text. Using the explicit amplitudes above, one obtains
\begin{align}
T &= \frac{A_0\,e^{-2Q_2 L}}{B_0 e^{-4Q_2 L} + B_1 e^{-2Q_2 L} + B_2},
\label{app:Graphene_T}
\\
R &= \frac{C_0\left(1-2e^{-2Q_2 L}\cos(2Q_1 L)+e^{-4Q_2 L}\right)}
{B_0 e^{-4Q_2 L} + B_1 e^{-2Q_2 L} + B_2}.
\label{app:Graphene_R}
\end{align}
The absorption is $A=1-T-R$. Note that the Lindblad evolution preserves $\mathrm{Tr}(\rho)$
 and $A$ quantifies flux missing from the coherent plane-wave channels.\\
The coefficients in Eqs.~(\ref{app:Graphene_T})--(\ref{app:Graphene_R}) are given as
\begin{widetext}
\begin{align}
A_0 &= 4 \cos^2(\phi)\left(\alpha_{-}^2 + 2 \alpha_{-}\alpha_{+}\cos(\theta_{-}+\theta_{+}) + \alpha_{+}^2\right),
\\
B_0 &= \left(\alpha_{-}^2 - 2\alpha_{-}\cos(\theta_{-}-\phi) + 1\right),
\\
B_1 &= 2\Big[
\alpha_{-}^2 \cos\!\big(2(Q_1 L + \phi)\big)
+ 2 \alpha_{-} \sin(2Q_1 L + \phi)\big(\alpha_{-} \alpha_{+} \sin\theta_{+} + \sin\theta_{-}\big)
\nonumber\\
&\qquad
- 2 \alpha_{+} \sin(2Q_1 L - \phi)\big(\alpha_{-} \alpha_{+} \sin\theta_{-} + \sin\theta_{+}\big)
+ \alpha_{+}^2 \cos(2Q_1 L - 2\phi)
\Big]
\nonumber\\
&\qquad
- 2 \cos(2Q_1 L)\Big(\alpha_{-}^2 \alpha_{+}^2 + 4 \alpha_{-}\alpha_{+}\sin\theta_{-}\sin\theta_{+} + 1\Big)
\Big(\alpha_{+}^2 - 2 \alpha_{+}\cos(\theta_{+}-\phi) + 1\Big),
\\
B_2 &= \left(\alpha_{-}^2 + 2\alpha_{-}\cos(\theta_{-}+\phi) + 1\right)
\left(\alpha_{+}^2 + 2\alpha_{+}\cos(\theta_{+}+\phi) + 1\right),
\\
C_0 &= \left(\alpha_{-}^2 + 2\alpha_{-}\cos(\theta_{-}+\phi) + 1\right)
\left(\alpha_{+}^2 - 2\alpha_{+}\cos(\theta_{+}-\phi) + 1\right).
\end{align}
\end{widetext}
\section{Over-the-barrier regime for graphene}
\label{app:tunneling}

Tunneling in the case of the over-the-barrier regime for graphene is shown in Fig. \ref{fig:phi_TAoverbarr} below. 
\begin{figure}[ht!]
\centering 
\includegraphics[scale=0.5]{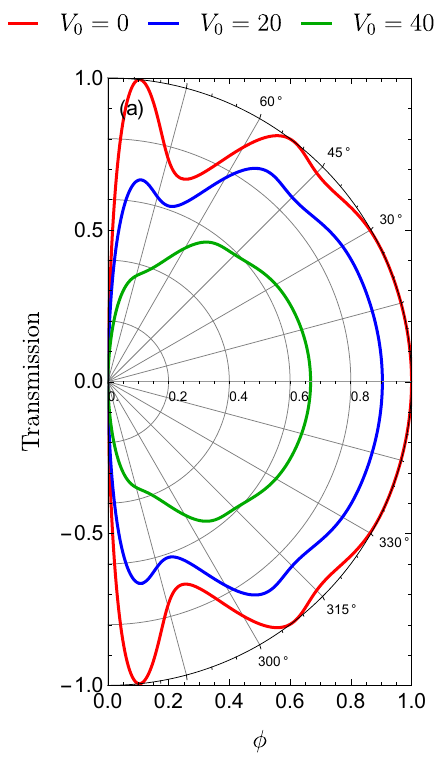}
\includegraphics[scale=0.5]{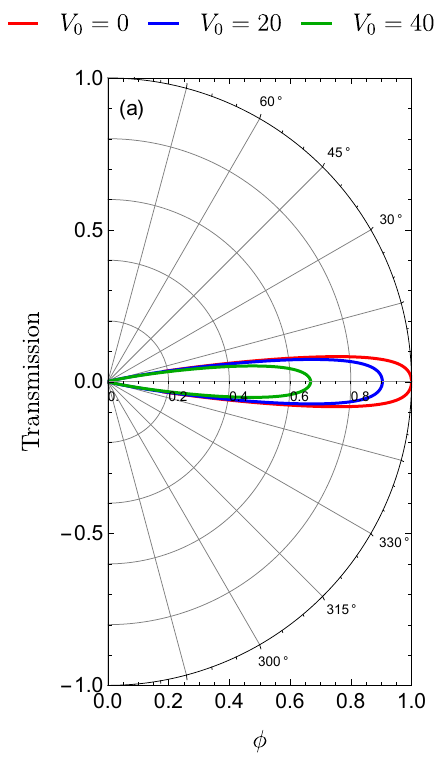}\\
\includegraphics[scale=0.5]{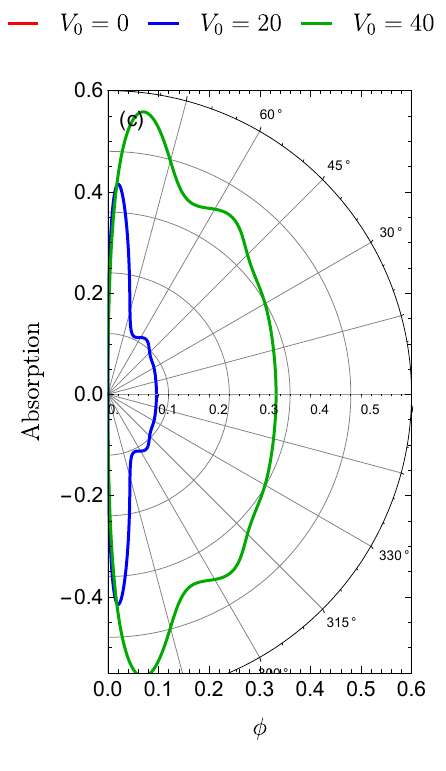}
\includegraphics[scale=0.5]{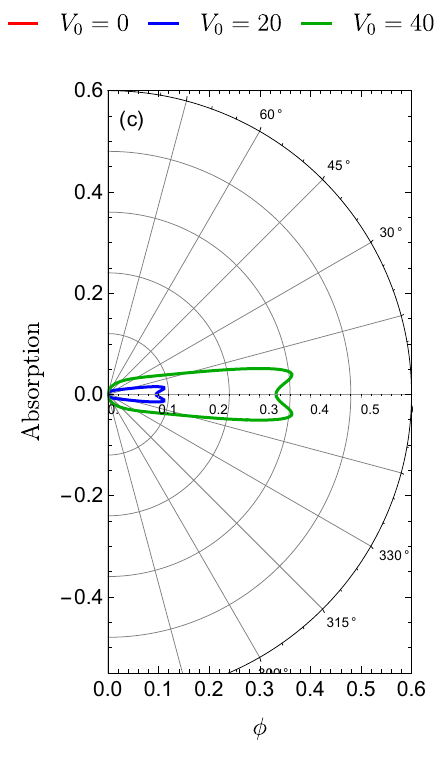}
        \label{fig:A_phi_sub}
 \caption{
    Angular dependence of transmission and absorption probabilities for tunneling through
    a noisy barrier in graphene. Panels (a,b) show the transmission $T(\phi)$ for $U_0=200$ and
    $285$~meV, and panels (c,d) show the corresponding absorption $A(\phi)$. The
    incoming energy is $\epsilon=300$~meV, the barrier width is $L=110$~nm. Curves correspond to the noise strength $V_0=0$, $20$, and
    $40$~meV.}
\label{fig:phi_TAoverbarr}
\end{figure}
We observe results similar to those in the tunneling regime.

\end{document}